\documentclass{aa}
\usepackage{txfonts}
\usepackage{graphicx}
\usepackage{natbib}

\begin{document}

\title{Fine-structure infrared lines from the Cassiopeia~A knots}

\author{D.\ Docenko\inst{1,2} \and R.A.\ Sunyaev\inst{1,3}}
\institute{Max Planck Institute for Astrophysics,
           Karl-Schwarzschild-Str.\ 1,
           85741 Garching, Germany
           \and
           Institute of Astronomy, University of Latvia,
           Rai\c{n}a bulv\={a}ris 19,
           Riga LV-1586, Latvia
           \and
           Space Research Institute, Russian Academy of Sciences,
           Profsoyuznaya 84/32,
           117997 Moscow, Russia
           }

\keywords{atomic processes -
          ISM: supernova remnants -
          infrared: ISM}


\abstract
{}
{Archival observations of infrared fine-structure lines of the young
Galactic supernova remnant Cassiopeia~A allow us to test existing
models and determine the physical parameters of various regions of
the fast-moving knots, which are metal-dominated clouds of material
ejected by the supernova explosion.
}
{The fluxes of far-infrared [\ion{O}{i}] and [\ion{O}{iii}] lines
are extracted from previously unpublished  archival ISO data.
The archival \emph{Spitzer} data are used to determine the fluxes of
the O, Ne, Si, S, Ar, and Fe ion fine-structure lines originating in
the fast-moving knots.
The ratios of these line fluxes are used as plasma diagnostics. We
also determine the infrared line flux ratios with respect to the
optical [\ion{O}{iii}] 5007~\AA\ line in the knots with previously
measured reddening.
Additionally, we analyze several optical and near-infrared
observations of the fast-moving knots to obtain clearer insight into
the post-shock photoionized region structure.
}
{We show that the infrared oxygen line flux predictions of all
existing theoretical models are correct only to within a factor of a
several. Comparison of the model predictions shows that to reproduce
the observations it is essential to include the effects of the
electron conductivity and dust.
Detailed analysis of the diagnostic line flux ratios of various ions
allows us to qualitatively confirm the general model of fast-moving
knot emission and determine observationally for the first time the
physical conditions in the photoionized region after the shock.
We infer from the [\ion{O}{iii}] line flux ratios that the pre-shock
cloud densities are higher than assumed in existing theoretical
models and most probably correspond to several hundred particles per
cm$^3$.
We also determine the Cas~A luminosity in the infrared continuum and
lines.
We show that accounting for the charge exchange processes in the
post-shock photoionized region allows us to reproduce most of the
relevant spectral line ratios even in the frame of a
single-temperature model of this region. We also estimate its plasma
parameters, thickness, and carbon abundance.
}
{}


\maketitle

\section{Introduction}
\label{SecIntro}

The young Galactic supernova remnant Cassiopeia~A (Cas~A) is
one of only a few objects where the inner composition of the
supernova progenitor may be studied directly, as the explosion
ejecta have not yet mixed with the surrounding medium.
In places where these ejecta are being crossed by the reverse shock
wave, they are observable as compact knots emitting optical
forbidden lines of ions of oxygen, sulfur, argon, and other
elements.
Because of their high density, these ejecta knots are not strongly
decelerated by the reverse shock and, as a result, have high proper
motions that was the reason to name them as the fast-moving knots
(FMKs).
Studies of \citet{Peimbert71} and \citet{CheKirCasA,CheKirAbund}
firmly established that the FMKs are an observational manifestation
of the supernova ejecta.

In FMKs, we therefore deal with a very unusual situation in
astrophysics, where oxygen comprises 80\%-90\% of all atoms in the
plasma, other components being mostly Ne, Si, S, Ar, and Fe. The
hydrogen and helium abundances are, in contrast, understood to be
negligibly low. Optical observations have shown that some knots
consist even of almost pure oxygen (so-called [\ion{O}{iii}]
filament, \citet{CheKirAbund}).

Theoretical interpretation of the optical observations has proven
that the plasma heated up to temperatures of several million degrees
is rapidly cooling because of its high density and the anomalous
chemical composition of the gas. At the same time, the ionization
degree lags behind the rapid cooling rate. This results in a unique
situation where atoms with strongly different ionization degrees
coexist at low temperatures from $10^4$~K to several hundreds
Kelvin.
Because of this, as we now show, the lines of highly-charged ions,
such as [\ion{Si}{x}] and [\ion{Ar}{v}], originate at temperatures
of about $10^4$~K and lower%
\footnote{%
The [\ion{Si}{x}] line has also been detected in spectra of several
active galactic nuclei \citep{AGNSiX,AGNNIRAtlas}, where it is
apparently formed in the photoionized region. Even so, the models
predict line-forming region temperatures of around $10^5$~K
\citep{SyFerland97}. }.

Traditionally, theoretical models of fast-moving knot emission were
compared only with optical line observations.
Since there is only a very limited number of diagnostic line ratios
in the optical spectrum, the FMK theoretical models are difficult to
both construct and compare with optical observations.
The most well-known models are those constructed by
\citet{Itoh81a,Itoh81b}, \citet{BorkowskiO}, and \citet{SD95}.

\cite{CasAORL} illustrated one way of constraining the theoretical
models using prospective observations of optical and near-infrared
recombination lines of highly-charged oxygen ions.

In this paper, we demonstrate how existing archival infrared
observations can discriminate between existing models and place
strong constraints on the construction of future ones.
We compare the model predictions of the ratios of far-infrared (FIR)
fine-structure oxygen line fluxes to the flux in the optical
[\ion{O}{iii}] 5007~\AA\ line with their observational values, and
show that all of the predictions are incorrect by a factor of
several for one or several flux ratios.
It should be remembered, however, that the theoretical models
predict the deviations of ionic abundances from their collisional
equilibrium values by several orders of magnitude. Therefore,
inconsistencies of the observed order suggest that some corrections
should be applied to the existing models, while the general picture
described in the models is entirely correct.

The infrared fine-structure line flux ratios of other element (in
addition to oxygen) ions are currently the most reliable tool
available for direct studies of the physical parameters in the FMKs.
We attempted to estimate these parameters using available
observational data.
As a result, parameters of the post-shock photoionized region were
estimated observationally for the first time, confirming
qualitatively the predictions of the theoretical models, but again
highlighting some quantitative differences.

The FIR lines will provide even more important probe in the future,
as several far-infrared observatories, such as the Herschel Space
Observatory and SOFIA, begin their operation.
These telescopes will provide far higher quality FIR data ofhigher
sensitivity and angular resolution, allowing us to constrain the FMK
models more tightly.
Results of this present study are also important to understand the
small-scale structure of other oxygen-rich supernova remnants, such
as Puppis~A, N132D, and G292+1.8.

\bigskip

The paper structure is the following. In the next section, we
describe briefly the existing theoretical models of the FMKs and
estimate oxygen infrared line fluxes in one of these models.
Section~\ref{SecIRObs} describes the archival infrared observations
of Cas~A and their analysis.
In Sect.~\ref{SecAnalysis}, we estimate the physical conditions and
abundances in different regions of FMKs from the observed line flux
ratios. We also discuss these results and compare them with values
predicted by the models.
In Sect.~\ref{SecPIR}, we analyze spectral line ratios to obtain a
consistent model of the post-shock photoionized region.
Section~\ref{SecIRL} is devoted to the predictions of the mid- and
far-infrared recombination line intensities.
Finally, in Sect.~\ref{SecConclusions} we present our conclusions.

\section{Theoretical models of fast-moving knots}
\label{SecModels}

The first detailed model of fast-moving knot emission, describing it
as originating from the passage of a shock through the pure oxygen
medium, was constructed by \citet{Itoh81a,Itoh81b}.
The predictions of his model H were compared with measurements of
the [\ion{O}{iii}] filament optical spectrum in the northern part of
the Cas~A supernova remnant (SNR) shell, and were found to reproduce
all four measured oxygen [\ion{O}{i}], [\ion{O}{ii}], and
[\ion{O}{iii}] optical line flux ratios to within a factor of two.
The infrared line ratios to the [\ion{O}{iii}] 5007~\AA\ line%
\footnote{%
The 5007~\AA\ line is three times brighter than the other [\ion{O}{iii}]
doublet component at 4959~\AA\ (e.g.,~\citet{Osterbrock06}).
Note that in some papers the line intensities are compared
to the sum of the doublet components.
}
for this theoretical model (denoted I-H) and other models described below
are given in Table~\ref{TabFIR}.

\begin{figure}
\begin{center}
\centerline{
    \rotatebox{270}{
        \includegraphics[height=0.95\linewidth]{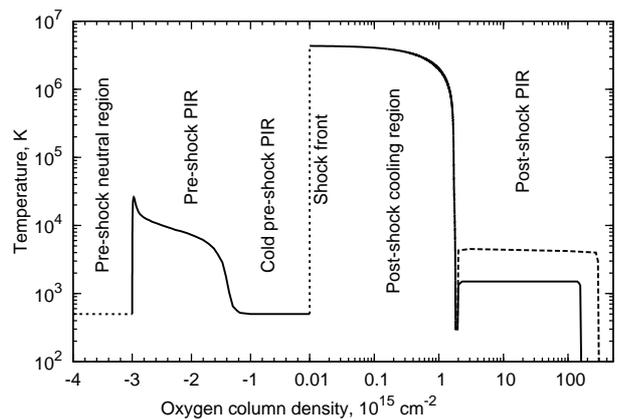}
                   }
           }
\caption{Schematic representation of the FMK temperature structure,
         induced by its interaction with the reverse shock.
         Negative column densities corresponding to the pre-shock regions
         are in linear scale;
         positive (post-shock) column densities are on a logarithmic scale.
         The post-shock photoionized region (PIR) parameters are given
         as derived in Sect.~\protect\ref{SecPIR} (solid line)
         and according to the \protect\cite{BorkowskiO} models
         (dashed line).
         Shock is moving to the left.
         Because of the FMK high density, the reverse shock in it has
         slowed down to about 200~km/s.
         }
   \label{FigStruct}
  \end{center}
\end{figure}

In this and other theoretical models, the rapidly cooling region
just after the shock front produces high ionizing radiation flux
that in its turn produces two photoionized regions (PIRs), before
and after the shock front (see Fig.~\ref{FigStruct}).

\begin{table*}
\caption{%
Far-infrared line flux ratios to the dereddened 5007~\AA\ line as derived
from different theoretical models and observations.
See text for details.
\label{TabFIR}
}
\begin{center}
\begin{tabular}{ll|lllll|l}
\hline\hline
  & $\lambda$, $\mu$m & \multicolumn{5}{c|}{$I/I(5007)$} &
                                                       $f_{\rm PIR}$ \\
& & I-H  & BS-F  & BS-DC & SD-200$^*$ & Observed$^{**}$ & SD-200$^*$\\
\hline
{}[\ion{O}{iv}] & 25.91 &
  0.0051& 5.0   & 0.53  &  8.1  & 0.20   & 0.995 \\
{}[\ion{O}{iii}]& 51.81 &
  0.031 & 1.34  & 0.72  &  0.67 & 0.25   & 0.999 \\
{}[\ion{O}{i}]  & 63.19 &
  1.30  & 170   & 12    &  ---  & 0.07   & ---   \\
{}[\ion{O}{iii}]& 88.36 &
  0.028 & 0.22  & 0.11  &  0.55 & 0.10   & 0.999 \\
{}[\ion{O}{i}]  & 145.5 &
  0.038 & 7.8   & 0.75  & ---   & $<$0.0024&---  \\
\hline
\multicolumn{7}{l}
     {$^*$ Derived in Sects.~\ref{SecIRSD} and \ref{SecLineI}.}\\
\multicolumn{7}{l}%
     {$^{**}$ Obtained in Sects.~\ref{SecISO} and \ref{SecSpitzerO}.}\\
\end{tabular}
\end{center}
\emph{Notes}. The last column lists contribution to the line flux in
the SD-200 model arising from the photoionized region (PIR) before
the shock. Derivation of the observed line flux ratio values is
described in Sect.~\protect\ref{SecIRObs}. The model-predicted
[\ion{O}{i}] lines are sensitive to the pre-shock ion number density
in the model, see Sect.~\ref{SecOI}.
\end{table*}

\citet{Itoh86} found that the model neutral oxygen optical line
intensities originating in the PIR after the shock front are much
too high compared to the observed spectra of the oxygen-rich
supernova remnant Cas~A and Puppis~A ejecta.
He suggested that emission from this region is damped because the
region itself is truncated due to some hydrodynamical phenomena.
This truncation affects essentially only neutral oxygen lines, as
all the other ions have already recombined in the dense post-shock
plasma before reaching the photoionized region.
This possibly ensures that the [\ion{O}{i}] lines are the least
reliable for direct comparison with model predictions, since all
existing models are one-dimensional.

Another group of theoretical FMK models was proposed by \citet{BorkowskiO}.
In contrast to the Itoh models, some of these accounted for electron
conductivity, which changed the line fluxes considerably.
One more difference is that the \citet{BorkowskiO} models do not
account for the emission from the PIR before the shock.
The models that most accurately describe the Cas~A FMK optical
spectra are F and DC (respectively, denoted as BS-F and BS-DC in
Table~\ref{TabFIR} and below).
We note that only the BS-DC model takes the electron conductivity
into account.

Both \citet{Itoh81a,Itoh81b} and \citet{BorkowskiO} assume that the
emission originates in the pure oxygen plasma.
However, it was shown by \citet{Dopita84} that inclusion
of other elements in the model
significantly changes the plasma thermal structure and emission,
especially in the cold PIRs, where a number
of fine-structure transitions in ions of other elements can effectively
cool plasma to temperatures of about one hundred Kelvin.
Following this line of reasoning, another group of the FMK models
was published by \citet{SD95}.
These models assume that the emission is produced by the interaction
of the dense cloud with the external shock, entering the cloud and
propagating through it.
The expected far-infrared line intensities are not published and we
estimate them in the Sects.~\ref{SecIRSD} and~\ref{SecLineI} for the
model with a 200~km/s shock speed and pre-shock ion number density
of 100~cm$^{-3}$ (denoted as SD-200 in Table~\ref{TabFIR} and
below), which most accurately reproduces the optical spectra of
Cas~A fast-moving knots, as demonstrated by \citet{SD95}.

Compared to the models of \citet{Itoh81a,Itoh81b} and
\citet{BorkowskiO}, the SD-200 model does not include the PIR after
the shock front, because it is constructed to describe the FMK
optical spectrum and the temperature of this post-shock PIR was
estimated to be too low to contribute significantly to the optical
line emission.

All described models nevertheless share many similar features (see
Fig.~\ref{FigStruct}).
In all of them, the plasma after the shock front passage rapidly
cools and, at temperatures of $(5-50)\times 10^3$~K, emits the
high-ionization lines observable in the visible and near-infrared
spectra \citep{CheKirCasA,CheKirAbund,HF96,CasANIR01}.
The thickness of this emitting layer is extremely small (about
$10^{10}$~cm for the pre-shock ion density of 100~cm$^{-3}$), but
because of the high electron density in the cooling material at
$T<10^5$~K (more than $10^5$~cm$^{-3}$), the emission measure is
enough to produce or contribute to bright emission lines of
highly-charged ionic species, such as [\ion{O}{iii}] lines near
5000~\AA.

In all of these models, the visible line emission of weakly-ionized
ionic species (e.g., [\ion{O}{i}] and [\ion{O}{ii}]) originates in
photoionized regions before and after the shock front.

Although all of the theoretical models describe optical spectra
relatively well, with discrepancies in relative intensities that do
not exceed factors of about two, predictions for the infrared lines
differ by several orders of magnitude (see Table~\ref{TabFIR}).
It is therefore clear that the models have a lack of constraining
power and more diagnostic information in the form of various line
flux ratios is needed to determine the true structure of the
fast-moving knots.
Part of this information may be obtained from the far-infrared line
archival observations, and we present this analysis in
Sect.~\ref{SecIRObs} below.

\subsection{Infrared lines from the SD-200 model}
\label{SecIRSD}

Because of a lack of published values, we had to make our own
estimates of the far-infrared line fluxes and their ratios with
respect to the [\ion{O}{iii}] 5007~\AA\ line in the SD-200 model.
Fortunately, all the data necessary for this calculation -- the
ionic abundances of all oxygen ions, the temperature and density
structure of the post-shock region, and the temperature structure of
the pre-shock photoionization front --
are known%
\footnote{%
We performed our own computations of the post-shock plasma
recombination and discovered that our oxygen ion distribution over
ionization stages is rather similar to that presented in the lower
panel of Fig.~3 of the \citet{SD95}, but only if the ion
spectroscopic symbols in that figure are increased by unity.
Therefore, we assume that Fig.~3 of \citet{SD95} has a misprint and
the ion spectroscopic symbols should be read, e.g., ``O~VII''
instead of ``O~VI'', ``O~VI'' instead of ``O~V'', etc.
From this, it follows that the SD-200 model values for the neutral
oxygen abundances are not known and we do not provide estimates of
the FIR [\ion{O}{i}] line intensities.}.

We note that the FMK structure calculations in the SD-200 model are
based on the plasma composition that differs slightly from the
composition inferred from the X-ray and optical observations.
Specifically, it is dominated by O (63\% by number) and Ne (29\%),
and contains minor amounts of C, Mg and Si. In contrast, the
observations imply that the majority of the currently bright X-ray
and optical plasma is oxygen (about 90\%), other abundant elements
being not only Ne, but also Si and S.

\begin{figure}
\begin{center}
\centerline{
    \rotatebox{270}{
        \includegraphics[height=0.95\linewidth]{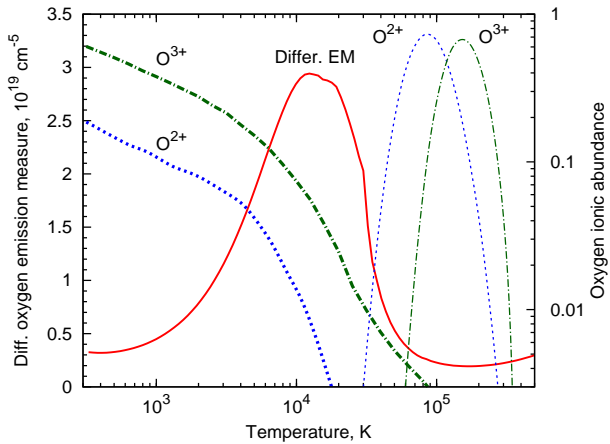}
                   }
           }
\caption{Comparison of ionic abundances of oxygen ions
         \element[2+]{O} and \element[3+]{O} in the collisional ionization
         equilibrium (thin curves) and the cooling post-shock region
         of the SD-200 model (thick curves of the same color).
         The differential oxygen emission measure per logarithmic temperature
         interval d$E_{\rm O}/$d\,ln$T_{\rm e}$
         is indicated by the red solid line.
         }
   \label{FigSDem}
  \end{center}
\end{figure}

In Fig.~\ref{FigSDem}, we demonstrate how the ionization state
distribution in the SD-200 model strongly differs from the
collisional ionization equilibrium (CIE, \citet{Mazzotta}) on
examples of the \element[2+]{O} and \element[3+]{O} ions producing
bright FIR lines.
It is seen that in the rapidly cooling plasma these ions are
abundant at significantly lower temperatures (down to 300~K and
below) than in the CIE because of very rapid plasma cooling by means
of line emission.

In the same figure, we present the oxygen ion emission measure
distribution over temperature d$E_{\rm O}/$d(ln$T_{\rm e}$) in the
SD-200 model, defined by Eq.~(\ref{EqEM}).
This distribution together with the line emissivity dependence on
temperature allows one to calculate the relative contributions to
the total line flux from different temperature intervals (see
Sect.~\ref{SecLineI}).

\subsection{Line flux computation}
\label{SecLineI}

The emissivity, cm$^3$/s, of a spectral line corresponding to transition
from level $u$ to level $l$ is defined as
\begin{equation}
\label{Eqeps}
\varepsilon(u,l) = n_{u} A_{ul} / n_{\rm e} n_{\rm i},
\end{equation}
where $n_{\rm e}$, $n_{\rm i}$, and $n_u$ are the number densities
of electrons, the corresponding ionic species, and ions in the state
$u$, respectively, and $A_{ul}$ is the spontaneous $u\to l$
transition rate.

For oxygen ions at temperatures above 1000~K, the line emissivities
$\varepsilon$ were obtained from the Chianti atomic
database~\citep{Chianti,Chianti5}.
For lower temperatures and other ions, we computed emissivities
using a standard approach, i.e., calculating the level populations
by solving a system of linear equations describing transitions
between the lowest ionic levels using electronic excitation
effective collision strength data from the literature and
spontaneous transition rates from both the \citet{NISTASD} database%
\footnote{Available online at \tt http://physics.nist.gov/asd3}
and MCHF/MCDHF collection\footnote{\tt http://atoms.vuse.vanderbilt.edu/},
and then applying Eq.~(\ref{Eqeps}) to derive the line emissivities.

The electronic collisional excitation effective collision strength
values were taken from \citet{SiIIFS91}, \citet{OIVFS92},
\citet{BlikeFS94}, \citet{OlikeFS94}, \citet{OIIIFS94},
\citet{SiSlikeFS95}, \citet{ArIIFS95}, \citet{FlikeFS98},
\citet{SIIIFS99}, \citet{NeVFS00}, \citet{NeIIFS01},
\citet{OIVFS06}, and \citet{FeIIFS07} and extrapolated to lower
temperatures by a constant, if needed.

Line fluxes $I(u,l)$, erg/cm$^2$/s, were computed by integrating
along the line of sight
\begin{equation}
\label{EqInt}
 I(u,l) = h\nu\frac{S}{4\pi R^2}
     \int \varepsilon\left(u,l; T_{\rm e}(r)\right)
        n_{\rm e}(r) n_{\rm i}(r) {\rm d}r,
\end{equation}
where $h\nu$ is the photon energy,
$R$ is the distance from  the observer to the emitting region
   (3.4~kpc, \citet{CasADist}), and
$S$ is the emitting region area.
From this expression, it is seen that the line emission mostly
originates in regions of the highest emission measure $n_{\rm e}
n_{\rm O} {\rm d}r$, ionic abundance $n_{\rm i}/n_{\rm O}$, and
emissivity $\varepsilon$ (here $n_{\rm O}$ is the total number
density of all oxygen ions).

The integral over distance in the post-shock cooling region can be
transformed into an integral over temperature by completing the
substitution
$$
 {\rm d}r = \frac{{\rm d}r}{{\rm d}t}
            \frac{{\rm d}t}{{\rm d}T_{\rm e}} {\rm d}T_{\rm e},
\qquad \frac{{\rm d}r}{{\rm d}t} =
   \upsilon_{\rm shock}\,\frac{n_{\rm 0,t}}{n_{\rm t}},
$$
where $\upsilon_{\rm shock}$ is the shock front speed, and
      $n_{\rm t}$ and $n_{\rm 0,t}$ are the total number densities
        of all ions in the plasma at a given point and before the
        shock, respectively
        (i.e., in the SD-200 model $n_{\rm 0,t}=100$~cm$^{-3}$).
It is assumed here that the photon heating is insignificant compared to
the energy losses.

The parameters of these equations, i.e., the cooling function, and
electron and ion densities as functions of temperature -- were taken
from the SD-200 model.

The derivative  ${\rm d}t/{{\rm d}T_{\rm e}}$ in the post-shock
cooling region may be obtained from the energy conservation law
written for one particle as
$$
  {\rm d} \left(
    \frac32 \frac{n_{\rm e}+n_{\rm t}}{n_{\rm t}} k_{\rm B}T_{\rm e}
  \right)
  =
  - n_{\rm e}\Lambda_{\rm N} {\rm d}t
  - p\,{\rm d}\left(\frac{1}{n_{\rm t}}\right),
$$
where $k_{\rm B}$ is the Boltzmann constant,
      $\Lambda_{\rm N}$ is the cooling function \citep{SD93}, and
      $p=(n_{\rm e}+n_{\rm t})k_{\rm B} T_{\rm e}$ is the gas pressure.
Dividing both sides by ${{\rm d}T_{\rm e}}$, we can now obtain the
required derivative from the expression
$$
  \frac32 k_{\rm B} T_{\rm e}
    \frac{{\rm d}(n_{\rm e}/n_{\rm t})}{{\rm d}T_{\rm e}}
  + \frac32 \left(\frac{n_{\rm e}}{n_{\rm t}}+1\right)k_{\rm B}
  =
  - n_{\rm e}\Lambda_{\rm N} \frac{{\rm d}t}{{\rm d}T_{\rm e}}
  + \left(\frac{n_{\rm e}}{n_{\rm t}}+1\right)
    \frac{kT_{\rm e}}{n_{\rm t}}
    \frac{{\rm d}n_{\rm t}}{{\rm d}T_{\rm e}}.
$$


We note that in the SD-200 model, the plasma pressure is not
constant in the post-shock cooling region (their Fig.~3) at low
temperatures because of the contribution of the magnetic field
pressure.

\begin{figure}
\begin{center}
\centerline{
    \rotatebox{270}{
        \includegraphics[height=0.95\linewidth]{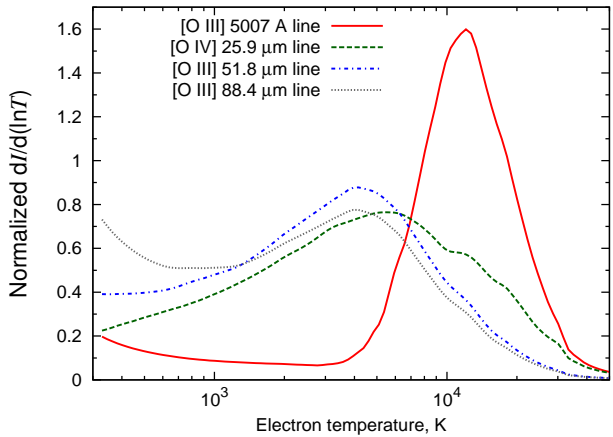}
                   }
           }
\caption{Differential flux contribution per logarithmic temperature
         interval for the optical [\ion{O}{iii}] 5007~\AA\ line and
         infrared oxygen lines in the cooling post-shock region
         according to the SD-200 model.
         Curves are normalized so that the
         area enclosed under each of them equals one.
         Low-temperature (below 4000~K) contribution
         to the optical [\ion{O}{iii}] line is due to recombination.
         }
   \label{FigSDdl}
  \end{center}
\end{figure}

For the purpose of qualitative analysis, we introduce the oxygen
differential emission measure per logarithmic temperature interval
\begin{equation}
\label{EqEM}
\frac{{\rm d}\,E_{\rm O}}{{\rm d}\, (\ln T_{\rm e})} =
  T_{\rm e} \, \frac{n_{\rm O} n_{\rm e} {\rm d}r}{{\rm d}T_{\rm e}}.
\end{equation}
It indicates contribution of a given logarithmic temperature
interval to the total emission measure, showing where most of the
line emission originates. Using this notation, we can express the
line flux given by Eq.~(\ref{EqInt}) as
$$
 I(u,l) =  h\nu\frac{S}{4\pi R^2}
     \int \varepsilon\left(u,l; T_{\rm e}\right)
          \frac{{\rm d}\,E_{\rm O}}
               {{\rm d}\, \ln T_{\rm e}}
          \;
          \frac{n_{\rm i}}{n_{\rm O}}
           \frac{{\rm d}T_{\rm e}}{T_{\rm e}}.
$$
Contributions from different temperature intervals to the emission
of the post-shock cooling region in the oxygen lines are illustrated
in Fig.~\ref{FigSDdl}, where we plot the normalized line fluxes per
logarithmic temperature interval.
It is seen that the post-shock contribution to the visible
[\ion{O}{iii}] line emission originates at temperatures of
$(5-20)\times10^3$~K, whereas the infrared lines are emitted mostly
at lower temperatures. A far broader range of temperatures also
contributes to the line emission because of lower excitation
energies.

In our analysis, we artificially stop integrating Eq.~(\ref{EqInt})
in the post-shock region when the plasma temperature drops below
300~K, since the SD-200 model does not contain data for lower
temperatures.
This should not influence the total FIR line intensities much, since
the model predicts that they are mainly emitted in the pre-shock
region.

In the pre-shock PIR, the ion density is constant and the
temperature dependence on time for the SD-200 model is known
explicitly. Therefore, Eq.~(\ref{EqInt}) was applied directly to
compute the line fluxes from this region.
The resulting line flux ratios are given in Table~\ref{TabFIR}.

\section{Archival observations of the FIR lines}
\label{SecIRObs}

The first search for the [\ion{O}{i}] 63.19~$\mu$m line was
performed by \citet{Harriet87}, who observed the northern part of
the Cas~A using the NASA Kuiper Airborne Observatory. This search
resulted in a 3-$\sigma$ upper limit to the reddening-corrected flux
ratio $I$(63.19~$\mu$m)/$I$(5007~\AA) of about 0.3, which tightly
constrained some theoretical models.
Later, the far-infrared lines of O, Ne, Si, S, and Ar ions were
detected by the Infrared Space Observatory
\citep[ISO,][]{ArendtCasA}.
We describe our analysis of some of these archival observations in
Sect.~\ref{SecISO}.
The \emph{Spitzer} Space Telescope recently spectrally mapped the
Cas~A supernova remnant~\citep{CasAIRAC,CasAIRS}. In
Sect.~\ref{SecSpitzer},
we determine the line ratios%
\footnote{Here and everywhere below the ``line ratio'' denotes
the ratio of the line fluxes in energy units.}
of O, Ne, Si, S, Ar, and Fe lines to the [\ion{O}{iii}] 5007~\AA\
line.

To determine these ratios, we compare the infrared observations
of ISO and \emph{Spitzer} with the optical images
of the Hubble Space Telescope.
The most accurate way of comparing these observations would be an
analysis of the properties of individual knots but, unfortunately,
the ISO data are of insufficient angular resolution (about
40\arcsec). The angular resolution of the \emph{Spitzer} data is
superior (2\arcsec$-$8\arcsec), but also too coarse to isolate the
contributions of \emph{individual} knots distinguishable within the
optical maps (0\farcs2$-$1\arcsec, corresponding to
$(3-15)\times10^{16}$~cm at 3.4~kpc).

As the optical observations show, the bright FMK lifetime is of the
order of 30 years \citep{KamperCasA}, which is determined by the
time needed by the shock to cross the cloud. After that, the knots
fade, but others gradually appear in the maps of the Cas~A.
Therefore, one may directly compare either observations performed
with a short time difference between them (as in the case of Hubble
Space Telescope observations compared with \emph{Spitzer} data), or
averaged over a large area (as in the case of the ISO data).

\subsection{ISO observations}
\label{SecISO}

The Infrared Space Observatory%
\footnote{%
Results in this section are based on observations with ISO, an ESA
project with instruments funded by ESA Member States (especially the
PI countries: France, Germany, the Netherlands, and the United
Kingdom) and with the participation of ISAS and NASA. } Long
Wavelength Spectrometer (ISO LWS) observations of the
\object{Cassiopeia A} supernova remnant were performed in 1996 and
1997.
Here we do not discuss observations at shorter wavelengths
($\lambda<40$~$\mu$m), since more recent data of significantly
higher angular resolution is available from the \emph{Spitzer}
observatory.

To determine the line fluxes, we used the ISO LWS calibrated data
from a highly-processed data product ``uniformly processed LWS L01
spectra'' available from the ISO data archive%
\footnote{
\tt http://www.iso.vilspa.esa.es/ida/
}.

\begin{figure}
\begin{center}
\centerline{
        \includegraphics[width=0.99\linewidth]{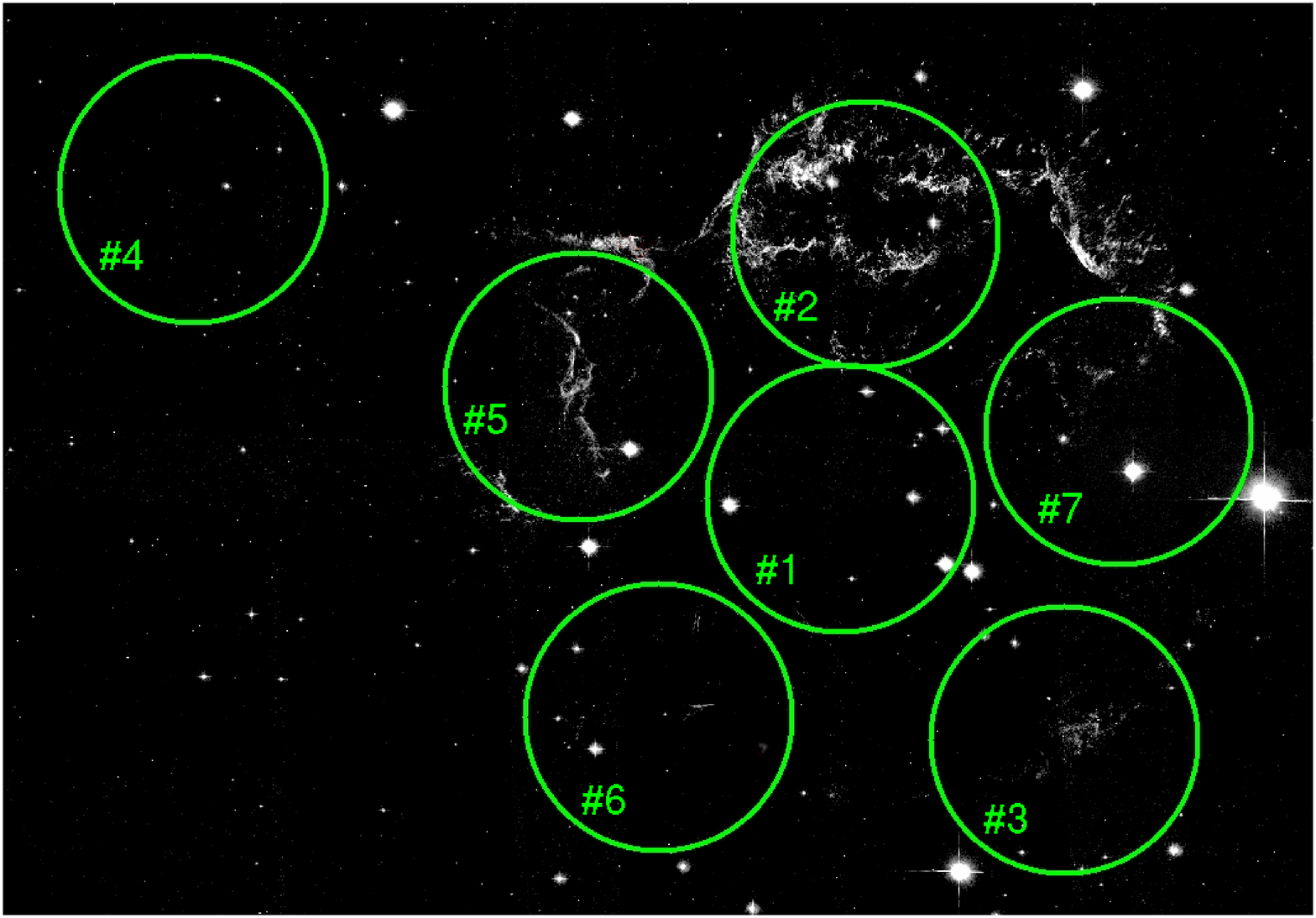}
           }
\caption{ISO LWS apertures (green circles)
         overlaid on the Hubble Space Telescope 2004 ACS image
         of Cas~A in F475W filter \citep{Fesen06}.
         North is upwards and East is to the left.
         }
   \label{FigISOobs}
  \end{center}
\end{figure}

There are seven LWS observations in L01 mode each covering the full
instrument spectral range between 43 and 170~$\mu$m with medium
spectral resolution ($\lambda/\Delta\lambda$ between 150 and 200).
The regions observed by ISO are shown in Fig.~\ref{FigISOobs}
overlaid on the Hubble Space Telescope Advanced Camera for Surveys
(HST ACS) Cas~A image in filter F475W, containing [\ion{O}{iii}]
4959 and 5007~\AA\ lines (Obs.ID 10286, observations completed in
December 2004, \citet{Fesen06}).

\begin{figure*}
\begin{center}
\centerline{
    \rotatebox{270}{
        \includegraphics[height=0.7\linewidth]{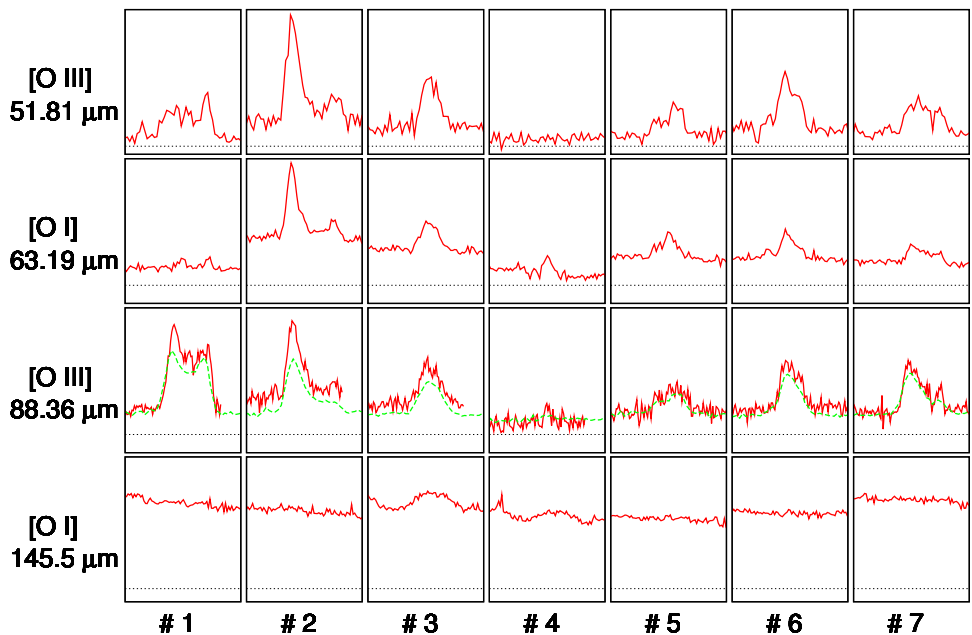}
                   }
           }
\caption{ISO LWS spectra of Cas~A at the observed locations (as
         shown in Fig.~\protect\ref{FigISOobs} and
         Table~\protect\ref{TabISOobs}) are shown in the regions of
         oxygen far-infrared lines discussed in the text.
         Continuum emission is not subtracted.
         Plotted are the averages over three scans weighted according
         to their uncertainties.
         Two curves for the 88.36~$\mu$m line correspond to measurements
         by two different ISO detectors.
         The short-dashed line denotes the zero level; maximum intensity
         corresponds to $1.7\times10^{-10}$, $0.7\times10^{-10}$,
         $0.7\times10^{-10}$, and $0.10\times10^{-10}$ erg/cm$^2$/$\mu$m
         for [\ion{O}{iii}] 51.81~$\mu$m, [\ion{O}{i}] 63.19~$\mu$m,
         [\ion{O}{iii}] 88.36~$\mu$m, and [\ion{O}{i}] 145.5~$\mu$m lines,
         respectively. Spectral ranges are centered on the rest-frame wavelengths
         of the respective spectral lines and include a velocity range
         of $\pm10^4$ km/s.
         }
   \label{FigISOspe}
  \end{center}
\end{figure*}

The ISO LWS observational data are summarized in
Table~\ref{TabISOobs} and the spectral cuts containing the discussed
lines are presented in Fig.~\ref{FigISOspe}.

\begin{table}
\caption{%
Summary of the ISO LWS archival observations of the Cassiopeia~A.
\label{TabISOobs}
}
\begin{center}
\begin{tabular}{lllll}
\hline\hline
ID        & Date & Exposure & R.A.     & Dec. \\
\hline
Cas A \#1 & 1996 Jun 24 & 1054 s  & 350.8656 & +58.8130\\
Cas A \#2 & 1996 Jun 24 & 1052 s  & 350.8614 & +58.8361\\
Cas A \#3 & 1996 Jun 24 & 1054 s  & 350.8279 & +58.7919\\
Cas A \#4 & 1996 Jun 24 & 1054 s  & 350.9748 & +58.8400\\
Cas A \#5 & 1997 Jun 09 & 1612 s  & 350.9097 & +58.8228\\
Cas A \#6 & 1997 Jun 09 & 1612 s  & 350.8963 & +58.7939\\
Cas A \#7 & 1997 Jun 02 & 1614 s  & 350.8186 & +58.8188\\
\hline
\end{tabular}
\end{center}
\end{table}

To compute the far-infrared spectral line fluxes, we subtracted the
background continuum flux and averaged over three scans of the same
detector, weighting the data according to its uncertainties.
In the case of the [\ion{O}{iii}] 88.36~$\mu$m line, it was possible
to estimate the systematic uncertainties by comparing the flux
values obtained from two neighboring detectors.
The flux differences are expected to have two causes: slightly
different apertures of the detectors and systematic errors due to
imperfect instrumental calibration \citep{ISOLWSHandbook}.
Our analysis showed that differences between the two detector
measurements do not exceed 10--15\%, when the line flux
determination accuracy is sufficiently high (see
Fig.~\ref{FigISOspe} for a visual comparison).

The background-subtracted FIR line fluxes from the observed regions
are presented in Table~\ref{TabISOflx}.
It is seen that the flux at the level of up to 10\% in the brightest
fields is observed in the 88.36~$\mu$m in the region \#4 situated
outside the main supernova remnant shell with no detected optical
[\ion{O}{iii}] emission and very little diffuse emission in other
optical bands.
We consider the flux from this region as originating in the
foreground and/or background of Cassiopeia~A and in our analysis
subtract it from the fluxes of the other regions.

\begin{table}
\caption{%
Far-infrared line fluxes from the ISO LWS observations of Cas~A.
Fluxes from the matched regions in the HST ACS F475W filter are also
given. The upper limits are at the $3\sigma$ level.
\label{TabISOflx} }
\begin{center}
\begin{tabular}{l|lllllll}
\hline\hline
     Line           &   \#1 &   \#2 & \#3 &   \#4 &   \#5 &   \#6 & \#7 \\
\hline
  \multicolumn{8}{c}{FIR line fluxes $F$, $10^{-11}$ erg/cm$^2$/s \large\strut}\\
{}[\ion{O}{iii}] 52 $\mu$m & 5.4 &   9.3 & 5.4 &$<$1.2 &   3.5 &   5.0 & 5.2 \\
{}[\ion{O}{i}] 63 $\mu$m   &$<$0.6&  2.6 & 1.2 &$<$1.2 &   1.0 &   1.0 & 0.9 \\
{}[\ion{O}{iii}] 88 $\mu$m & 7.5 &   4.2 & 2.7 &   0.5 &   1.9 &   3.2 & 3.4 \\
{}[\ion{O}{i}] 145 $\mu$m  &$<$0.1&$<$0.09&0.3 & 0.1   &$<$0.1 &$<$0.1 &$<$0.1 \\
\hline
  \multicolumn{8}{c}{ [\ion{O}{iii}] 4959 + 5007 \AA\ line flux $F$,
                      $10^{-13}$ erg/cm$^2$/s   \large\strut}\\
                    &$<$0.1 & 25    & 2.0 &$<$0.1& 6.5   & 0.5   & 2.9\\
\hline
\end{tabular}
\end{center}
\emph{Note}. ISO region positions are shown in Fig.~\ref{FigISOobs}.
Regions \#3, \#6 and \#7 have higher interstellar absorption column
\citep{BiegingOH}.
\end{table}

\begin{table}
\caption{%
Far-infrared line flux ratios to the optical 5007~\AA\ line
$I/I$(5007~\AA)
corrected for the optical line reddening.
\label{TabISOlirat}
}
\begin{center}
\begin{tabular}{l|lllll}
\hline\hline
     Line                  & \#2     &  \#5   \\
\hline
{}[\ion{O}{iii}] 51.81 $\mu$m & 0.25    &   0.36  \\
{}[\ion{O}{i}] 63.19 $\mu$m   & 0.07    &   0.10  \\
{}[\ion{O}{iii}] 88.36 $\mu$m & 0.10    &   0.14  \\
{}[\ion{O}{i}] 145.5 $\mu$m  &$<$0.0024&$<$0.01  \\
\hline
\end{tabular}
\end{center}
\emph{Note}. The 5007~\AA\ line contribution of 75\% to the total
FMK emission in the F475W filter is assumed. Optical line
attenuation due to reddening of 200~times is assumed in both the ISO
LWS regions, see text. FIR line attenuation is neglected.
\end{table}

In the same table, we indicate the total fast-moving knot optical
flux for the matched regions in the F475W filter of the discussed
HST ACS data.
The diffuse emission in this filter is assumed to come only from the
[\ion{O}{iii}] 4959 and 5007~\AA\ lines.

From these data, we calculated the line ratios to the [\ion{O}{iii}]
5007~\AA\ line, given in Table~\ref{TabISOlirat}.
Assuming that the FIR line emission observed by ISO originates in
the same spatial regions as the optical line emission, these flux
ratios are characteristic of the knot emission and may be used for
tests of the theoretical models.
If significant large-scale diffuse emission in the FIR lines is
present, as observed in the [\ion{S}{iii}] and [\ion{O}{iv}] lines
by Spitzer observations (see Sect.~\ref{SecSpitzer}), then the
determined line ratios should be interpreted as upper limits to the
values characteristic of the knots.

In Table~\ref{TabISOlirat}, we provide the values corrected for the
interstellar absorption of the optical line using an average value
from \citet{HF96}.
This level of reddening diminishes the observed visible
[\ion{O}{iii}] doublet intensities by a factor of about 200
($A_V\approx5.0$) and is characteristic of region \#2 and, possibly,
region \#5 \citep{BiegingOH}.

For southern and western regions (\#3, \#6 and \#7) the reddening is
known to be much higher because of absorption in the molecular
clouds with peak $A_V$'s reaching 10--15 \citep{BiegingOH}
corresponding to the optical line attenuation of up to factors of
$10^4$ and more.
Therefore, we do not provide dereddened line ratios in these
regions.
In our further analysis in Sects.~\ref{SecAnalysis}
and~\ref{SecComp}, we use only the data from the region \#2.

There is an eight-year time difference between ISO and HST observations.
During this time, which is about 30\% of a bright FMK lifetime
\citep{KamperCasA},
some optical knots may have disappeared, and new ones may have appeared.
However, averaging over a large number of optical knots contained in
any ISO LWS field of view at least partially cancels changes induced
by the brightening or fading of individual knots.

\subsection{\emph{Spitzer} Space Telescope observations}
\label{SecSpitzer}

The \emph{Spitzer} Space Telescope spectrally mapped the
Cassiopeia~A in the infrared range using its Infrared Spectrograph
(IRS) producing low-resolution spectra from 5 to 38~$\mu$m
\citep{CasAIRAC,CasAIRS} and detected in this spectral band the
fine-structure lines of [\ion{O}{iv}], [\ion{Ne}{ii}],
[\ion{Ne}{iii}], [\ion{Ne}{v}], [\ion{Si}{ii}], [\ion{S}{iii}],
[\ion{S}{iv}], [\ion{Ar}{ii}], [\ion{Ar}{iii}], [\ion{Ar}{v}],
[\ion{Fe}{ii}], etc.

We used basic calibrated data of these publicly available
\emph{Spitzer} observations (Program 3310) to construct the data
cube over the entire remnant with the CUBISM software
\citep{cubism}.
The background regions were determined from the 8~$\mu$m
MIPS images of the supernova remnant \citep{CasAMIPS}.

The line flux maps were then produced using the parabolic
approximation for the background continuum emission.
This allowed the maps to be constructed reliably even for
the weak lines in the regions of strong continuum dust emission.

Since the \emph{Spitzer} spectral maps and the HST images were
obtained almost simultaneously with a time difference of only one
month, we were able to make a direct pixel-by-pixel comparison of
the infrared line map with the optical image.
For this purpose, we processed the HST images by removing the stars,
 convolving with the point spread function (PSF)
   of a given \emph{Spitzer} spectral map
   computed as described in Appendix~\ref{AppPSF},
 and regridding the HST image to match pixels
   to the IRS module containing each infrared line.
These matched HST images were produced and compared to each of the
infrared lines.

\subsubsection{The oxygen [\ion{O}{iv}] 25.91~$\mu$m line}
\label{SecSpitzerO}

\begin{figure}
\begin{center}
\centerline{
        \includegraphics[width=0.95\linewidth]{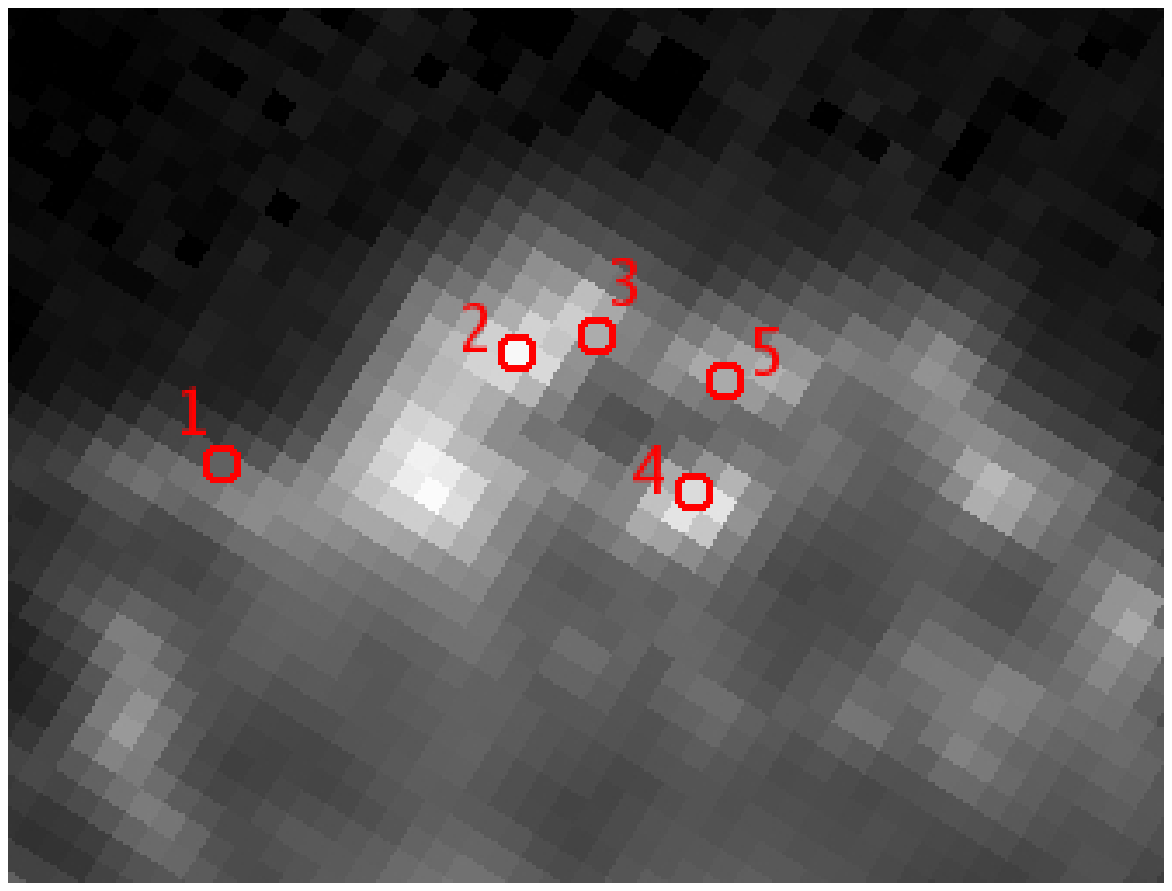}
           }
\centerline{
        \includegraphics[width=0.95\linewidth]{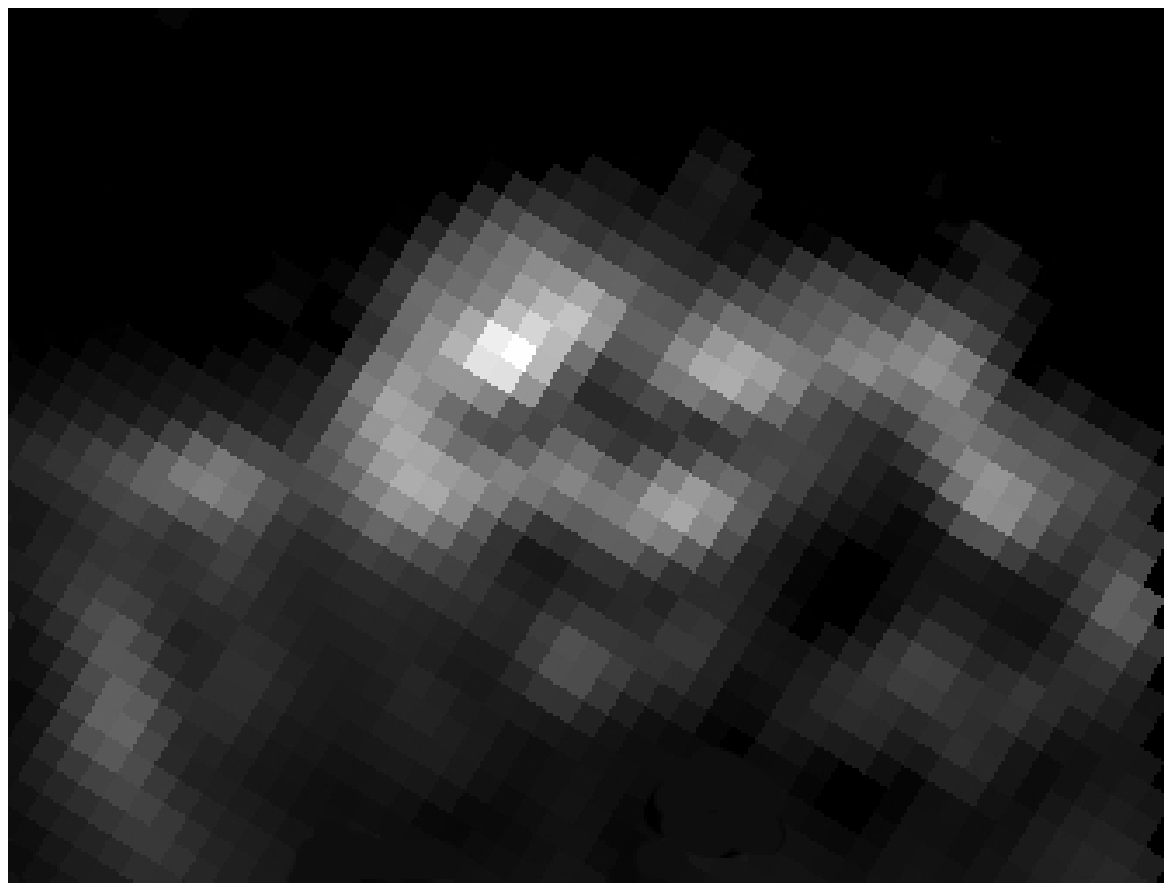}
           }
\caption{The \emph{Spitzer} Space Telescope map of the
         northern part of the Cas~A optical shell rich in FMKs
         in the [\ion{O}{iv}] line at 25.91~$\mu$m (top)
         compared with the HST image in F475W filter
         matched in resolution and in pixels (bottom).
         The maps are centered on R.A. of 350.8614 deg and
         declination of 58.8361 deg and show the region of $2.5\times3.5$ arcminutes.
         The red circles denote regions with reddening measured
         by \protect\citet{HF96}.
         Square root scale in intensity is used to enhance weak
         features. Diffuse infrared [\ion{O}{iv}] emission
         from optically-dark regions is clearly visible.
         }
   \label{FigIRmap}
  \end{center}
\end{figure}

The [\ion{O}{iv}] 25.91~$\mu$m spectral line map of the northern
Cas~A shell region obtained from \emph{Spitzer} observations is
presented in Fig.~\ref{FigIRmap}.
Alongside, we show the HST image in F475W filter processed
as described above.
The northern region is of primary importance as it has a
well-measured and almost constant interstellar absorption level.

It can be readily seen from the brightest features in
Fig.~\ref{FigIRmap} that there is a difference between the Cas~A
morphologies for the optical and infrared lines. The difference may
have several reasons: variations in the interstellar reddening
\citep{HF96}, emitting plasma chemical composition
\citep{CheKirAbund}, and FMK pre-shock density distribution over the
remnant.

\begin{figure}
\begin{center}
\centerline{
    \rotatebox{270}{
        \includegraphics[height=0.95\linewidth]{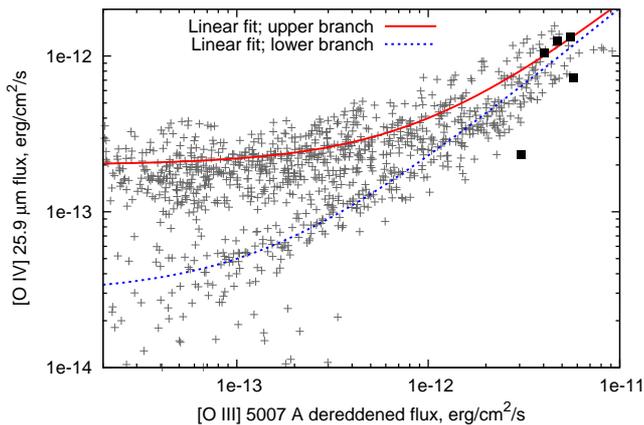}
                   }
           }
\caption{Pixel-by-pixel comparison of the dereddened 5007~\AA\ line
         flux with the infrared [\ion{O}{iv}] line map.
         Fluxes are given per detector pixel,
         equal to $(5\farcs08)^2$ for the IRS LL module containing the
         infrared oxygen line.
         Two linear fits, both with the same slope of 0.20,
         are shown with lines.
         Filled squares denote FMKs from \protect\citet{HF96}
         with measured reddening.
         }
   \label{FigIRvsHST}
  \end{center}
\end{figure}

In Fig.~\ref{FigIRvsHST}, we present the resulting pixel-by-pixel
scatter plot showing interrelation between the [\ion{O}{iv}]
infrared and [\ion{O}{iii}] optical line fluxes.
From this comparison of the maps, we derive the average line ratio
of the optical and infrared lines, which corresponds to an average
reddening, FMK density, and chemical composition.

In this analysis, ``average'' reddening was assumed to diminish the
observed optical [\ion{O}{iii}] line emission by a factor of 200,
which is an average of the \citet{HF96} measurements.
As may be inferred from their paper, small-scale variations in the
interstellar reddening in this region change the optical line
attenuation within a factor of two. The observed scatter of the line
ratio seen in Fig.~\ref{FigIRvsHST} is of similar magnitude.

In the same figure, we also present data for five FMKs with
reddening measured by \citet{HF96}. Their positions are shown in
Fig.~\ref{FigIRvsHST}. In these cases, the reddening correction was
made according to their measurement results. Four out of five points
lie very close to the best-fit lines, showing that a considerable
part of the FMKs have similar excitation conditions.
A remaining point (FMK~1) has relatively weaker [\ion{O}{iv}]
emission, showing that the physical conditions there are different.
This region (Filament~1 of \citet{BaaMin54}) indeed has a
composition that differs from that of the other regions
\citep{CheKirAbund}.

In Fig.~\ref{FigIRmap}, infrared emission in the [\ion{O}{iv}] line
that originates in the central Cas~A region with no optical
nebulosity (see also \citet{CasAIRS}) is also noticeable.
In Fig.~\ref{FigIRvsHST}, this infrared-only emission shows up on
the left as horizontal branches, corresponding to an approximately
constant background surface brightness. Two of these branches are
apparent, probably corresponding to two different column densities
of the \element[3+]{O} ion.
Pixels forming the lower branch are situated just outside the
optical shell, where the optical surface brightness is already
fading, but the infrared line emission remains present.

The upper branch corresponds to the diffuse emission in the central
part of the supernova remnant.
Corresponding points in the HST image are partially outside the
Fig.~\ref{FigIRvsHST} plot range, but the infrared line surface
brightness remains on the level of
$(1-3)\times10^{-13}$~erg/cm$^2$/s/pix.

This infrared-only line emission is consistent with the existence of
extended cold photoionized regions invisible in optical and X-ray
emission close to the center of the supernova remnant, possibly
connected with the thermal radio absorption observed in the same
regions \citep{CasAMHz,TraceyPhD}.

We note that the [\ion{O}{iv}] infrared line at 25.91~$\mu$m
overlaps with the [\ion{Fe}{ii}] line at 25.99~$\mu$m, but the
contribution of the latter can be estimated by comparison with the
[\ion{Fe}{ii}] 17.94~$\mu$m line map.
We assume that the [\ion{Fe}{ii}] 25.99/17.94~$\mu$m line ratio is
constant over the remnant near the reverse shock. Then, we select a
bright region of Cas~A with a low value of the line flux ratio
$$
\frac{I(\mbox{[\ion{O}{iv}] 25.91~$\mu$m + [\ion{Fe}{ii}]
25.99~$\mu$m})} %
{I(\mbox{[\ion{Fe}{ii}] 17.94~$\mu$m})} \approx 5,
$$
and assume it to be the upper bound to the [\ion{Fe}{ii}]
25.99/17.94~$\mu$m line ratio later used in Sect.~\ref{SecPIR}.

This bound is used to infer that the upper limit to the 25.99~$\mu$m
[Fe~II] line contribution to the total flux of the observed line
near 26~$\mu$m is about $20-40$\% in all regions of the Cas~A used
in our analysis (i.e., the \citet{HF96} FMKs).
%

\subsubsection{Other infrared lines in the \emph{Spitzer} spectral maps}
\label{SecSpitzerOthers}

Similar pixel-by-pixel comparisons with the optical image were also
performed for other bright spectral line flux maps.
In Table~\ref{TabAIR}, we provide the derived flux ratios to the
optical [\ion{O}{iii}] 5007~\AA\ line in five \citet{HF96} FMKs and
the average values of these ratios in the northern part of the Cas~A
shell. The last digit in brackets in the ``average'' column denote
the observed 2-$\sigma$ scatter around the average line flux ratio
in units of the last digit.
This scatter represents the true variations in the line flux ratios
due to both chemical composition variations \citep{CheKirAbund} and
variations in physical excitation conditions (e.g., different
pre-shock density in different regions of Cas~A).

\begin{table*}
\caption{%
Infrared line fluxes $I$ and their ratios $I/I$(5007~\AA)
  to the dereddened optical
  [\ion{O}{iii}] 5007~\AA\ line from the \emph{Spitzer} data.
The column ``average'' contains the ratio values averaged over the
northern part of Cas~A and their observed scatter, and other columns
contain data for the \citet{HF96} FMKs. \label{TabAIR} }
\begin{center}
\begin{tabular}{l|lllll|l|lllll}
\hline\hline
Line
    & \multicolumn{5}{c|}{Line flux $I$, $10^{-13}$ erg/cm$^2$/s/pixel}
    & \multicolumn{5}{c }{Flux ratio $I/I$(5007~\AA)} \\
    & FMK 1 &FMK 2 &FMK 3 &FMK 4 &FMK 5
    & Average
    & FMK 1 &FMK 2 &FMK 3 &FMK 4 &FMK 5\\
\hline
{}[\ion{Ar}{ii}]  6.985 $\mu$m
    & 3.8  & 14   & 17   & 5.6  & 11
    & 0.5(1)
    &0.41  &0.75  &1.15  &0.46  &0.79 \\
{}[\ion{Ar}{iii}] 8.991 $\mu$m
    & 0.46 & 2.0  & 1.7  & 0.80 & 1.5
    & 0.08(3)
    &0.05  &0.13  &0.12  &0.07  &0.12 \\
{}[\ion{S}{iv}]   10.51 $\mu$m
    & 0.24 & 1.6  & 1.4  & 0.67 & 0.91
    & 0.07(2)
    &0.025 &0.11  &0.10  &0.06  &0.07 \\
{}[\ion{Ne}{ii}]  12.81 $\mu$m
    & 0.31 & 0.92 & 0.89 & 1.0  & 0.30
    & 0.10(8)
    &0.04  &0.07  &0.07  &0.08  &0.025\\
{}[\ion{Ne}{v}]   14.32 $\mu$m
    & 0.071& 0.20 & 0.18 & 0.10 & 0.14
    & 0.012(7)
    &0.009 &0.012 &0.012 &0.009 &0.014\\
{}[\ion{Ne}{iii}] 15.56 $\mu$m
    & 0.29 & 2.2  & 1.2  & 3.0  & 0.47
    & 0.05(3)
    &0.009 &0.03  &0.03  &0.05  &0.006\\
{}[\ion{Fe}{ii}]  17.94 $\mu$m
    & 0.16 & 0.90 & 1.3  & 1.8  & 0.87
    & 0.03(2)
    &0.005 &0.015 &0.03  &0.04  &0.013\\
{}[\ion{S}{iii}]  18.71 $\mu$m
    & 0.82 & 5.0  & 5.2  & 2.5  & 4.2
    & 0.08(2)
    &0.025 &0.08  &0.15  &0.05  &0.06 \\
{}[\ion{O}{iv}]   25.91 $\mu$m
    & 2.3  & 13   & 11   & 12   & 7.2
    & 0.20(5)
    &0.08  &0.24  &0.26  &0.26  &0.13 \\
{}[\ion{S}{iii}]  33.48 $\mu$m
    & 0.28 & 1.9  & 2.3  & 0.76 & 0.6
    & 0.04(1)
    &0.011 &0.05  &0.06  &0.015 &0.015\\
{}[\ion{Si}{ii}]  34.81 $\mu$m
    & 1.0  & 3.7  & 3.9  & 1.7  & 1.7
    & 0.07(3)
    &0.04  &0.10  &0.11  &0.03  &0.04 \\
\hline
\end{tabular}
\end{center}
\emph{Notes}. The table contains data for the strongest lines only.
The regions are numbered as in \citet{HF96} with their positions
shown in Fig.~\protect\ref{FigIRvsHST}.
The ``average'' column assumes single optical line reddening
correction of 200~times. Note that the pixel sizes are different for
the IRS LL (5\farcs08 or $6.06\times10^{-10}$~sr,
$\lambda>15$~$\mu$m) and SL (1\farcs85 or $8.04\times10^{-11}$~sr,
$\lambda<15$~$\mu$m) modules.
\end{table*}

From the table, it can be seen that different line flux ratios have
different scatter around their average values.
A small scatter in e.g., the Ar, Si, and S line ratios to the
5007~\AA\ line implies that the Cas~A morphology in these lines is
similar to the optical one.
A larger scatter, present in e.g., Ne and Fe line ratios to the
5007~\AA\ line corresponds to different morphologies of the infrared
spectral maps, as also noted by \citet{CasAIRS}.

The obtained line ratios may also be used as tests of the
theoretical models and are utilized in the next section to estimate
the average abundances and physical conditions in the bright Cas~A
knots.

From Table~\ref{TabAIR}, it is seen that the full energy flux in the
infrared lines detected by \emph{Spitzer} exceeds one of the
brightest optical line
 -- [\ion{O}{iii}] 5007~\AA\ --
and is of the same order as the total power in all
optical lines summed up.

\subsection{Comparison of the model predictions with observations}
\label{SecComp}

As the models provide predictions only for the oxygen line relative
intensities, we cannot use observations of lines of other ions to
directly compare with the models.

Comparing observed far-infrared line ratios with theoretical model
results (see Table~\ref{TabFIR}), we see that the
\citet{Itoh81a,Itoh81b} models underestimate the [\ion{O}{iii}] and
[\ion{O}{iv}] line fluxes by factors of 3--40, as a result of having
a far lower average ionization degree after the shock passage than
in other models.
In contrast, the \citet{BorkowskiO} model BS-F and the \citet{SD95}
model SD-200 that do not account for electron conductivity, both
overpredict the emission in the [\ion{O}{iv}] line by factors of
$20-40$.

The remaining model (BS-DC of \citet{BorkowskiO}) predicts the
infrared line fluxes to within a factor of several, except for the
neutral oxygen lines, but this problem is at least partially
resolved by increasing the pre-shock density (see
Sect.~\ref{SecOI}).
This strongly suggests that taking into account the electron
conductivity is essential for a model when attempting to reproduce
the observations.

The [\ion{O}{iii}] far-infrared line ratio
$I$(51.81~$\mu$m)/$I$(88.36~$\mu$m) in the \citet{BorkowskiO} models
is significantly higher than observed (6 versus 2.5).
This discrepancy arises because they do not include emission from
the pre-shock photoionized region.
In the SD-200 model, in contrast, the pre-shock region dominates the
[\ion{O}{iii}] infrared line emission and diminishes this infrared line
flux ratio to 1.2.
Therefore, to reproduce the observed line ratio the pre-shock and
post-shock contributions should be comparable (see also
Table~\ref{TabLinRat}).

The model I-H provides the most accurate description of the neutral
oxygen FIR line ratio. In this model, most of the [\ion{O}{i}]
emission originates in the post-shock PIR.
The line ratio corresponds to the high-density limit, but the predicted
line intensities are about 20 times stronger than observed.
As mentioned in Sect.~\ref{SecOI}, this discrepancy may be reduced
if we assume higher pre-shock atom number density than in the model
(for example, 300~cm$^{-3}$ instead of 30~cm$^{-3}$), as the line
emissivities decrease, but both the ionization parameter and the
column density of the PIR remain the same.

\subsection{Energetics of the infrared emission}
\label{SecEnergy}

From the analysis of the \emph{Spitzer} archival observations, it is
straightforward to determine that in the wavelength range from 5 to
35~$\mu$m the total Cas~A luminosity is $3.3\times10^{37}$~erg/s
(the luminosity values are computed assuming a distance to Cas~A of
3.4~kpc; \citet{CasADist}). The total spectral line luminosity in
the same range is found to be one order of magnitude lower,
contributing $3.0\times10^{36}$~erg/s.

For the longer wavelengths between 35~$\mu$m and 1~mm, we estimated
the total Cas~A luminosity using data published by \citet{CasAMIPS}
to be about $1.5\times10^{37}$~erg/s. The line contribution to this
value in the narrower ISO LWS range between 40 and 200~$\mu$m is
about $9\times10^{35}$~erg/s, as follows from Table~\ref{TabISOflx}.
Therefore, the total Cas~A dust continuum and line luminosities in
the infrared range between 5~$\mu$m and 1~mm are $4.7\times10^{37}$
and $3.9\times10^{36}$~erg/s, respectively.

It is interesting to compare these values with the X-ray and radio
luminosities.
In radio, the total luminosity at wavelengths longer than 1~mm is
about $1.0\times10^{35}$~erg/s as calculated directly from the
spectrum provided by \citet{CasAFlux} and \citet{CasAMIPS}, i.e.,
much lower than in the infrared.

It is not so straightforward to determine the total Cas~A luminosity
in X-rays because of interstellar and internal X-ray absorption.
Various estimates result in values corrected for the absorption
of $(1-5)\times10^{37}$~erg/s \citep{CasAROSAT,Zombeck3ed},
but are dependent on the assumed spectral model and absorbing material
column density.

It can be seen from this comparison that the infrared continuum
emission is as an important radiative energy loss mechanism from the
supernova remnant as the X-rays.
%

It is easy to notice that the infrared line-to-continuum emission
ratio in the \emph{Spitzer} range is variable over the supernova
remnant, as indicated also by the spectra provided by
\citet{CasAIRS}, with a typical value of 0.10. The highest values of
this ratio of about 0.22 are characteristic of the central Cas~A
region. In these regions, the energy losses in the line emission are
relatively twice as high as average, although, still significantly
below the losses in the infrared continuum.

\section{Physical conditions and abundances in the FMKs}
\label{SecAnalysis}

The measured line fluxes may be used directly to determine the
ionic abundances%
\footnote{%
In this paper, by abundance we denote a ratio of number density of
given ions or atoms to the number density of all oxygen ions
(i.e., $n/n_{\rm O}$).
},
if the model of the emitting region (temperature
and density distribution) is known.
Sicne the models do not reproduce observations precisely enough (see
Table~\ref{TabFIR}), we cannot assume that this distribution is
known for the FMKs.

Thus, the line fluxes cannot be used in such a direct fashion and
the inferred ionic abundances will depend on the assumed physical
conditions.
Fortunately, emissivities of some of the fine-structure lines are
only weakly dependent on the electron densities and temperatures
accross a rather wide range, allowing us to use these lines for the
abundance determination even without detailed knowledge of the
line-emitting region properties.

An analysis of the measured flux ratios of lines of the same ion is
a more powerful method. These line ratios may be used directly to
constrain the temperature and density of the emitting region,
provided that the emitting region is uniform and physically the same
for both lines defining the ratio.

However, these assumptions are not fulfilled in the fast-moving
knots of most line pairs and we cannot obtain emitting region
parameters from, e.g., a comparison of optical and infrared lines of
[\ion{O}{iii}].
Even the density-sensitive ratios of two infrared lines should be
interpreted carefully, as from the theoretical models it follows
that their emission may originate both in the pre-shock and
post-shock regions that have comparable emission measures of the
order of $10^{19}$~cm$^{-5}$ and temperatures of the order of
$(0.5-2)\times10^4$~K, but densities that differ by a factor of a
thousand or more.

More specifically, the pre-shock photoionized region has a low
electron density (of the order of $100-250$~cm$^{-3}$ according to
the SD-200 model) and produces lines with approximately low-density
limit line ratio, and the post-shock cooling region ($n_{\rm
e}\approx 10^6$~cm$^{-3}$) produces lines having high-density limit
line ratio (i.e., the corresponding level populations are determined
by collisional processes).
Therefore, instead of determining the density, some of the FIR line
ratios determine the contributions of the high- and low-density
regions to the total line emission.

In the following subsections, we analyze the line flux ratios to
infer the physical conditions and the line fluxes to estimate the
ionic and elemental abundances.

\subsection{Information obtained from the same ion line flux ratios}
\label{SecLinRat}

\begin{table*}
\caption{%
Information from the same ion line flux ratios measured in the FMKs.
See text for details.
\label{TabLinRat}
}
\begin{center}
\begin{tabular}{lll|l|ll|llll}
\hline\hline
  & & & Flux & \multicolumn{2}{c|}{Single region} &
                \multicolumn{4}{c}{Two regions}\\
Spectrum  & line $a$ & line $b$ & ratio &
           $T_{\rm e}$, K $^{*}$ & $n_{\rm e}$, cm$^{-3}$ &
                 $R_1$ & $R_2$ & $f_{{\rm PIR}, a}$ & $f_{{\rm PIR}, b}$ \\
\hline
{}[\ion{O}{i}]   & 63.19~$\mu$m   & 145.5~$\mu$m & $>30$ &
           $<100$              & any \\
& & & &    $200 - 1\times10^4$ & $>2\times10^5$ \\
{}[\ion{O}{iii}] & 51.81~$\mu$m & 88.36~$\mu$m & 2.5 &
           $300-1.5\times10^4$ & $500-1000$ &
                 1.1   & 9.7   & 0.40  &  0.85\\
{}[\ion{O}{iii}] & 88.36~$\mu$m & 5007~\AA & 0.10 &
           $1.0\times10^4$     & 800 & & & 1.0$^{*}$ & 1.0$^{*}$ \\
& & & &    $1.5\times10^4$     & 200 & & & 1.0$^{*}$ & 1.0$^{*}$ \\
{}[\ion{Ne}{iii}] & 36.01~$\mu$m & 15.56~$\mu$m & $<0.08$ &
           any & any \\
{}[\ion{Ne}{v}]   & 24.32~$\mu$m & 14.32~$\mu$m & $<0.2$  &
           $<3\times10^4$      & $>3\times10^4$ \\
{}[\ion{S}{iii}]  & 33.48~$\mu$m & 18.71~$\mu$m & 0.5    &
           $600 - 3\times10^4$ & $1500-3000$ &
           1.8 & 0.09 & 0.85 & 0.25 \\
{}[\ion{Ar}{iii}] & 21.83~$\mu$m & 8.991~$\mu$m  & $<0.10$ &
           any & any\\
{}[\ion{Ar}{v}]   & 7.914~$\mu$m & 13.07~$\mu$m  & $1.6$ &
           $1500-3\times10^4$ & $(4-10)\times10^4$ &
           1.0 & 6.7 & 0.5 & 0.9\\
{}[\ion{Fe}{ii}]$^{***}$  &1.3209~$\mu$m &1.2946~$\mu$m & 0.75   &
           any  & $>2\times10^5$ \\
{}[\ion{Fe}{ii}]$^{***}$  &1.3209~$\mu$m & 17.94~$\mu$m & 0.024  &
           1700$-$3000$^{**}$  & any \\
 & & & &   1700$-$1800$^{**}$  & $>1\times10^5$ $^{*}$\\
\hline
\end{tabular}
\end{center}
$^{*}$ Assumed value \\
$^{**}$ Derived value\\
$^{***}$ The [\ion{Fe}{ii}] line flux ratios
         are discussed in Sect.~\ref{SecDiscPre}.
\end{table*}

In the infrared range probed by the \emph{Spitzer} and the ISO
observatories, there are seven line pairs that can be used for
plasma diagnostics, and have at least one of lines defining the pair
detected.
These line pairs correspond to transitions between three of the
lowest fine structure states $^3P_{0,1,2}$ in the ground $p^2$ and
$p^4$ electron configurations.
Of these seven diagnostic line pairs, three have both lines detected
([\ion{O}{iii}], [\ion{S}{iii}], and [\ion{Ar}{v}] lines) and four
have detections of just one of the lines ([\ion{O}{i}],
[\ion{Ne}{iii}], [\ion{Ne}{v}], and [\ion{Ar}{iii}] lines).
The measured values of the flux ratios and the inferred physical
parameters are summarized in Table~\ref{TabLinRat}.

\subsubsection{[\ion{O}{iii}] FIR line flux ratio}

The measured ratios of the far-infrared [\ion{O}{iii}] lines
$I$(51.81~$\mu$m)/$I$(88.36~$\mu$m) in different ISO spectra lie
between 0.7 and 2.5 (see Table~\ref{TabISOflx}) with some indication
of a correlation between the line ratio and the optical line
intensity
(a correlation is also observed for the [\ion{S}{iii}] line ratio
map compiled from the \emph{Spitzer} data).

In the approximation of two emitting regions (pre-shock and
post-shock), we use the following expression to determine the
pre-shock PIR contribution to the total line $b$ flux from the
measured $a$ and $b$ line flux ratio $R=I_a/I_b$ (see
Appendix~\ref{AppfPR} for its derivation):
\begin{equation}
\label{EqfPR}
 f_{{\rm PIR}, b} \equiv \frac{I_{{\rm PIR}, b}}{I_{{\rm tot}, b}}
                 = \frac{R_2-R}{R_2-R_1},
\end{equation}
where $I_{{\rm PIR}, b}$ and $I_{{\rm tot}, b}$ denote the
photoionized region and total line $b$ fluxes, respectively, and
$R_1$ and $R_2$ denote theoretical line ratio values in the pre- and
post-shock regions.

The [\ion{O}{iii}] line ratio of 2.5 observed for the ISO LWS region
\#2 with many optical knots may thus be achieved only if 85\% of the
88.36~$\mu$m line and 40\% of the 51.81~$\mu$m line originate in the
pre-shock region.
The corresponding theoretical line ratios in this case are $R_1=1.1$
and $R_2=9.7$ (see Table~\ref{TabLinRat}).

As shown by~\citet{SD95}, the 5007~\AA\ line mostly originates in
the pre-shock region. We determined that this is also the case for
the 88.36~$\mu$m line. This allows us to take one further step and
use their flux ratio of $I$(88.36~$\mu$m)/$I$(5007~\AA$)\approx0.10$
to estimate the pre-shock physical conditions in the region emitting
these [\ion{O}{iii}] lines.

\begin{figure}
\begin{center}
\centerline{
    \rotatebox{270}{
        \includegraphics[height=0.95\linewidth]{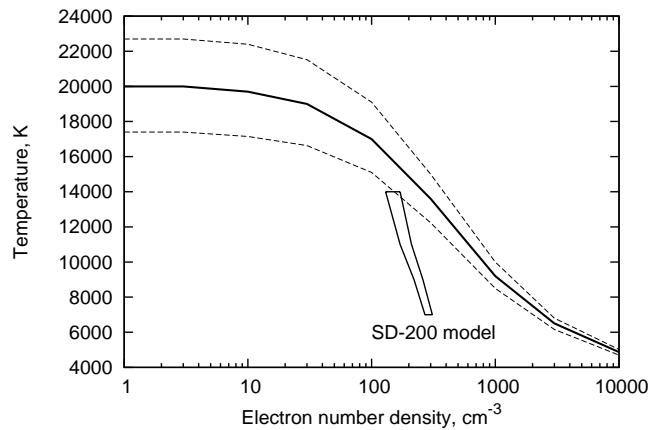}
                   }
           }
\caption{The constraints on the pre-shock region emitting the [\ion{O}{iii}]
         lines from the average $I$(88.36~$\mu$m)/$I$(5007~\AA) line flux ratio
         (solid line). The dashed lines correspond to the constraints
         if the line flux ratio is changed by 20\%.
         The contour denotes the temperature-density relation
                 in the region of the SD-200 model most intensely emitting
                 [\ion{O}{iii}] lines.
         }
   \label{FigO3}
  \end{center}
\end{figure}

This line ratio is sensitive to both density and temperature and in
a single-temperature emission region model corresponds to $n_{\rm
e}$ of several hundreds~cm$^{-3}$ for temperatures
$(1.0-1.5)\times10^4$~K (see Table~\ref{TabLinRat}). These values
are very similar to, but somewhat different from, ones given by the
SD-200 model for which $T_{\rm e}\le 1.1\times10^4$~K and $n_{\rm e}
\approx 200$~cm$^{-3}$ in the region that most intensely emits both
[\ion{O}{iii}] lines.
The resulting observational constraint on the pre-shock line
emission region is indicated in Fig.~\ref{FigO3}.

\subsubsection{[\ion{O}{i}] FIR line flux ratio}
\label{SecOI}

It is more difficult to use the [\ion{O}{i}] line ratio for the
quantitative analysis, because only one of the lines is detected.
From ISO observations of region \#2, we obtain the 3$\sigma$ lower
limit to the $I$(63.19~$\mu$m)/$I$(145.5~$\mu$m) line ratio of about
30.

Some information may still be obtained about this lower limit: it
corresponds to either extremely low temperatures below $100$~K or
high electron densities above $10^5$~cm$^{-3}$ (see
Table~\ref{TabLinRat}).

According to the theoretical models, there is essentially no neutral
oxygen in the post-shock cooling region because of slow
recombination processes. The models also predict the rapid
ionization of neutral oxygen in the pre-shock PIR with a very low
line ratio $I$(63.19~$\mu$m)/$I$(5007~\AA) much less than 0.01 (the
observed value is 0.07, see Table~\ref{TabISOlirat}). The pre-shock
region in which the [\ion{O}{i}] line originates, is also excluded
from the observed value of the $I$(63.19~$\mu$m)/$I$(145.5~$\mu$m)
line ratio.

From this, we conclude that the far-infrared [\ion{O}{i}] line
emission in the FMKs originates in the dense photoionized region
after the shock.

All the models predict values of this line flux ratio with respect
to the 5007~\AA\ line to be much higher than observed (see
Table~\ref{TabFIR}).
Part of this discrepancy may be explained by the low model values
for the pre-shock oxygen densities.
For example, the \citet{BorkowskiO} models assume a very low
pre-shock oxygen density of 1~cm$^{-3}$. Taking it higher by two
orders of magnitude will decrease the post-shock 63.19~$\mu$m line
intensity significantly. However, this is still not enough to
reconcile the model predictions with the observations.
The most probable reason for the low observed
$I$(63.19~$\mu$m)/$I$(5007~\AA) line ratio is a combination of the
post-shock region truncation and high pre-shock density, as
discussed below in Sect.~\ref{SecPIRSize}.

It should be noted that based on ISO data we cannot exclude part of
the 63.19~$\mu$m line originating in a cold ($T<100$~K or so)
diffuse medium outside the optical knots (although, see
considerations in Sect.~\ref{SecPIRSize}). This possibility may be
fully assessed only by high angular resolution observations, which
only new generation observatories such as \emph{Herschel} are able
to perform.

\subsubsection{The [\ion{Ne}{iii}], [\ion{Ar}{iii}], [\ion{Ne}{v}],
                [\ion{Ar}{v}] and [\ion{S}{iii}] line flux ratios}

All these lines are detected in the \emph{Spitzer} data and, because
of its much higher angular resolution, provide much more information
for the analysis.

The upper limits to the [\ion{Ne}{iii}] and [\ion{Ar}{iii}] line
ratios (see Table~\ref{TabLinRat} for details on these and other
line ratios) do not constrain the plasma parameters much, since both
of them correspond to a wide range of densities and temperatures
\citep{RubinRatios}.

Limits to the [\ion{Ne}{v}] line ratio also do not provide strong
constraints partly because of spectral fringing near 20~$\mu$m in
the IRS spectra \citep{CasAIRAC} that increases the effective noise
level.
Nevertheless, the limit to the [\ion{Ne}{v}] 24.32~$\mu$m line
implies that the [\ion{Ne}{v}] emission mostly or entirely
originates in the post-shock cooling region.

As for the case of [\ion{O}{iii}], we derived the fraction of the
fine-structure line emission originating in the pre-shock region for
[\ion{S}{iii}] and [\ion{Ar}{v}] lines.
Most of the [\ion{S}{iii}] line at 33.48~$\mu$m originates in the
pre-shock region, which we use to estimate the S abundance in
Sect.~\ref{SecLinInt}.

Existence of \element[4+]{Ar} in the pre-shock region (see
Table~\ref{TabLinRat}) is understandable, given its relatively low
\element[3+]{Ar} ionization potential of 60~eV, comparable to the
ionization potential of \element[2+]{O} (55~eV).
Therefore, similar amounts of O$^{2+}$ and O$^{3+}$ in the pre-shock
photoionized region in the SD-200 model give us a reason to deduce
that Ar$^{3+}$ and Ar$^{4+}$ ions also have similar number
densities.
We note that \citet{CheKirAbund} demonstrated that the visible
[\ion{Ar}{iv}] lines originate in the low-density environment, which
we associate with a pre-shock PIR.

\subsection{Abundances from the flux ratios to the 5007~\AA\ line}
\label{SecLinInt}

In the general case, ionic and elemental abundances derived from the
individual line intensities are strongly dependent on the
theoretical model of the fast-moving knots.
This model dependence may be minimized if the spectral lines are
known to be emitted in the pre-shock or the post-shock region only.

Some dependence on the underlying model still remains, but the model
features used in our analysis are relatively robust. For example,
the peak of the post-shock plasma emission measure around $10^4$~K
is determined by the cooling rate that slows down at $2\times10^4$~K
as resonance ultraviolet lines of various ions stop being
effectively excited.
The low-temperature limit of the line-emitting region is computed from
the upper transition level excitation energy,
which is a simple but reliable estimate.

The ionic abundances in the pre-shock region are stronger dependent
on the underlying model, since we estimate them for all elements
from average ionization potential of the oxygen ions in the SD-200
model.
Thus, the pre-shock abundances may be incorrect by up to a factor of
a several.

As follows from the theoretical models, spectral lines of singly ionized
species (e.g., [\ion{Ar}{ii}], [\ion{Si}{ii}], etc.)
arise mostly in the photoionized regions, whereas ones of highly-ionized
species (e.g., [\ion{Ne}{v}], [\ion{Mg}{v}], etc.) arise mostly
in the post-shock cooling regions.
The only exception is the [\ion{Ar}{v}] line, but the pre-shock
region contribution to its lines have already been estimated above.

Below we discuss these two line groups separately and summarize the
results in Tables~\ref{TabLinesPost} and~\ref{TabLinesPre}.

\begin{table*}
\caption{%
Estimated FMK post-shock cooling region  ionic abundances
 $n_{\rm ion}/n_{\rm O}$. See text for details.
\label{TabLinesPost}
}
\begin{center}
\begin{tabular}{ll|lll|ll}
\hline\hline
Line  & $I/I$(5007~\AA)
      & $T_{\rm e}$, K & $\varepsilon$, cm$^3$/s
      & $\Delta t$, $10^3$ s & $n_{\rm ion}/n_{\rm O}$
                      & $n_{\rm ion}/n_{\rm O}$ (model)$^*$
{\large\strut}\\
\hline
{}[\ion{Si}{x}]  1.430 $\mu$m  & 0.0016
      & $2\times10^4$ & $2.9\times10^{-8}$ &
                  40           & $7\times10^{-5}$ & $5\times10^{-4}$ \\
{}[\ion{Si}{vi}] 1.964 $\mu$m  & 0.004
      & $1\times10^4$ & $3.1\times10^{-9}$ &
                  40           & 0.002            & 0.01   \\
{}[\ion{Mg}{v}]  5.608 $\mu$m  & $<0.003$
      & $6\times10^3$ & $9.0\times10^{-9}$ &
                  30           & $<$0.002         & $4\times10^{-4}$  \\
{}[\ion{Ar}{v}]  7.914  $\mu$m & 0.005$^{**}$
      & $1\times10^4$ & $1.3\times10^{-8}$ &
                  60           & $8\times10^{-4}$ & $5\times10^{-5}$  \\
{}[\ion{Ne}{v}] 14.32  $\mu$m  & 0.015
      & $1\times10^4$ & $2.4\times10^{-9}$ &
                  50           & 0.06             & 0.002  \\
\hline
\end{tabular}
\end{center}
$^*$ Estimated based on SD-200 model.
The elemental abundances assumed to deduce the model ionic
abundances with respect to oxygen are given in Table~\ref{TabAbund}.\\
$^{**}$ This line is partly arising in the pre-shock region.
As follows from Table~\ref{TabLinRat},
the post-shock region contribution is
$I/I$(5007~\AA$)\approx0.0025$.
\end{table*}

\begin{table*}
\caption{%
Estimated FMK pre-shock ionic abundances $n_{\rm ion}/n_{\rm O}$
computed assuming that the lines originate only in the pre-shock
PIR. See text for details. \label{TabLinesPre} }
\begin{center}
\begin{tabular}{ll|lllll|ll}
\hline\hline
Line  & $I/I$(5007~\AA)
      & $T_{\rm e}$, K & $n_{\rm e}$, cm$^{-3}$ & $\varepsilon$, cm$^3$/s
      & IP, eV & $\Delta t$, s & $n_{\rm ion}/n$(O)
                                        & $n_{\rm element}$/$n$(O) $^{**}$
{\large\strut}\\
\hline
{}[\ion{Ar}{ii}]  6.985 $\mu$m & 0.5
      & $2.0\times10^4$ & 100 & $4\times10^{-8}$
      &  27.6 & $5\times10^6$ & 1.0$^*$ & 0.005 \\
{}[\ion{Ar}{v}]  13.07 $\mu$m & 0.003
      & 7000            & 250 & $8\times10^{-7}$
      &  75.1 & $5\times10^7$ & $8\times10^{-5}$ & 0.005\\
{}[\ion{Ne}{ii}] 12.81 $\mu$m & 0.10
      & $1.5\times10^4$ & 150 & $5\times10^{-8}$
      &  41.0 & $1.4\times10^7$ & 0.06  & 0.02  \\
{}[\ion{Fe}{ii}] 17.94 $\mu$m & 0.03
      & $2.0\times10^4$ &  50 & $1.1\times10^{-8}$
      &  16.2 & $9\times10^5$ & 5.5$^*$ & 0.005  \\
{}[\ion{S}{iii}] 33.48 $\mu$m & 0.04
      & $1.5\times10^4$ & 150 & $3.0\times10^{-7}$
      &  34.8 & $1\times10^7$ & 0.015   & 0.05  \\
{}[\ion{Si}{ii}] 34.81 $\mu$m & 0.07
      & $2.0\times10^4$ &  50 & $1.5\times10^{-7}$
      &  16.4 & $1\times10^6$ & 0.7$^*$ & 0.05  \\
{}[\ion{O}{iii}] 88.36 $\mu$m & 0.10
      & $1.0\times10^4$ & 200 & $4.0\times10^{-8}$
      &  54.9 & $3\times10^7$ & 1.1     & 1.0\\
\hline
\end{tabular}
\end{center}
$^*$ Shown to arises mostly in the post-shock photoionized region;
     see Sect.~\ref{SecPIR}.\\
$^{**}$ Atomic number densities with respect to oxygen, see
Table~\ref{TabAbund}.
\end{table*}

Because of extreme chemical inhomogeneities in the Cas~A supernova
remnant, one should not expect abundances in X-ray emitting
low-density ejecta to be the same as in much denser
optically-emitting knots. In our studies of infrared lines from
dense knots, the abundances determined from optical spectra should
be more reliable. However, optical abundance precision is not
expected to be superior to a factor of two, because of the very
approximate physical models used in the data analysis. Optical data
are also not available for Si and are especially uncertain for Fe
and Mg \citep{CheKirCasA}.

For our analysis, we summarized the available data on abundances in
the Cas~A ejecta (see Table~\ref{TabAbund}) and also used some
``average'' values. However, as noted above and clearly seen also
from Table~\ref{TabAbund}, the observed abundances vary dramatically
from place to place.

\begin{table*}
\caption{%
Summary of elemental abundance $n_{\rm element}/n_{\rm O}$ values in
Cas~A ejecta from various sources. The ``average'' data used in this
paper are shown in the last column. \label{TabAbund} }
\begin{center}
\begin{tabular}{l|lllll|l}
\hline
       & CK78 & CK79        & Wil02 & LH03 & Laz06 & Used in this paper \\
\hline
Method & Optical & Optical   & X-ray & X-ray & X-ray & --- \\
\hline
Ne/O   & 0.025 & $<$0.013    & 0.02  & ---          & ---          & 0.02  \\
Mg/O   &$<$0.04& $<$0.005    & 0.004 & ---          & 0.005$-$0.03 & 0.005 \\
Si/O   & ---   & ---         & 0.025 & 0.003$-$0.015& 0.008$-$0.06 & 0.05  \\
S/O    & 0.06  & 0.002$-$0.7 & 0.014 & ---          & 0.006$-$0.05 & 0.05  \\
Ar/O   & 0.004 & 0.003$-$0.1 & 0.0035& ---          & ---          & 0.005 \\
Fe/O   & 0.003 & $<$0.001 & 0.006$-$0.02& 0.002$-$0.01 & ---       & 0.005 \\
\hline
\end{tabular}
\end{center}
Data source abbreviations: \citet{CheKirCasA} -- CK78,
\citet{CheKirAbund} -- CK79, \citet{CasAXMM2002} -- Wil02,
\citet{LamingHwang03} -- LH03, \citet{CasAChandra} -- Laz06.
\end{table*}

\subsubsection{High-ionization ionic species
     from the post-shock cooling region}

An easy estimate of the plasma composition may be made if the
problem is simplified by assuming that emission originates in a
single layer with given temperature and electron density. In this
case one can estimate relative number densities of two types of ions
from their line fluxes.
The line flux ratio is then expressed from  Eq.~(\ref{EqInt}) as
\begin{equation}
\label{EqPoSLR}
\frac{I_1}{I_2} =
   \frac{\nu_1\; \varepsilon_1\;
         n_{\rm ion,1}\; \Delta t_1}
        {\nu_2\; \varepsilon_2\;
         n_{\rm ion,2}\; \Delta t_2},
\end{equation}
where $n_{\rm ion}$ is ionic number density,
$\Delta t$ is the line emission duration in the
post-shock cooling plasma element,
and it is assumed that the electron densities are the same for
regions emitting both lines.
%
The line emissivities entering Eq.~(\ref{EqPoSLR}) are computed as
described in Sect.~\ref{SecLineI}.

We note that the application of Eq.~(\ref{EqPoSLR}) is possible in
the SD-200 model because both the ion and electron density remain
approximately constant at $T\la 10^4$~K. This, combined with the
relatively weak dependence of the fine-structure line emissivity on
temperature, allows us to use this simple estimate without
introducing errors of more than a factor of two or so.

Equation~(\ref{EqPoSLR}) may be inverted to determine the relative
ionic abundances from the observed line ratios.
In it, we always assume the second line in the ratio to be the
[\ion{O}{iii}] 5007~\AA\ line.

In the following, we use the SD-200 model with the post-shock
\element[2+]{O} abundance in the line emitting region of
$n({\rm O^{2+}})/n({\rm O})\approx1$\%.
We also take into account that only about 2\%
of the total 5007~\AA\ line emission originates in the post-shock
plasma according to the SD-200 model.
This fraction is uncertain by up to a factor of about two, thus
introducing a systematic error of similar magnitude in all
post-shock ionic abundance determinations.

There are no published results about the theoretically predicted
abundances of ions other than oxygen in the post-shock cooling
region of the FMKs.
To estimate them, we utilized the electron and ion densities and the
cooling function from \citet{SD95} and traced the ion ionization and
recombination processes in the post-shock cooling phase
using atomic data compiled by Dima Verner%
\footnote{\tt http://www.pa.uky.edu/\~{ }verner/atom.html}
for the Cloudy code~\citep{Cloudy}.

Another source of uncertainty in the ionic abundance determination
using our method are the model-dependent values of $\Delta t$,
which were also computed by tracking the plasma cooling and
recombination.
However, calculations show that for different ions and lines the
$\Delta t$ values differ by no more than approximately a factor of
two, as determined by the general emission measure behavior.
The resulting model abundances are compared with the observational
results in Table~\ref{TabLinesPost}.

It is seen that the abundances inferred from the SD-200 model seem to
be generally consistent with the results of observations, showing that
the model in general correctly represents the temperature and density
evolution in the post-shock cooling region.
The [\ion{Ar}{v}] and [\ion{Ne}{v}] lines are the only ones showing
significant differences from the model expectations, although the
derived ionic abundances do not significantly exceed the
corresponding elemental abundances.

We note that the two highest ionization fine-structure lines of
[\ion{Si}{x}] and [\ion{Si}{vi}] observed from the fast-moving knots
have not been observed by \emph{Spitzer}, but detected in the
near-infrared spectra by \citet{CasANIR01}. We determined their line
flux ratios to the [\ion{O}{iii}] 5007~\AA\ line using reddening
measurements from the same regions by \citet{HF96}.

\subsubsection{Low-ionization ionic species from the pre-shock PIR}
\label{SecLinesPre}

The post-shock plasma recombination is significantly slower than the
cooling and, even at $T_{\rm e}\approx300$~K, the post-shock cooling
plasma still mostly contains multiply-charged ions, resulting in
essentially no infrared line emission from singly ionized atoms at
wavelengths shorter than several tens of $\mu$m.

In this section, we neglect any contribution to the line fluxes from
the post-shock \emph{photoionized} region.
Its possible existence and physical conditions are discussed in
Sect.~\ref{SecPIR}.

We can then write Eq.~(\ref{EqInt}) in a form similar to
Eq.~(\ref{EqPoSLR}) for the pre-shock region given by
\begin{equation}
\label{EqPrSLR}
\frac{I_1}{I_2} =
   \frac{\nu_1\; \varepsilon_1\; n_{\rm e, 1}\;
         n_{\rm ion,1}\; \Delta t_1}
        {\nu_2\; \varepsilon_2\; n_{\rm e, 2}\;
         n_{\rm ion,2}\; \Delta t_2}.
\end{equation}
The equation also contains the electron density ratio, because ions
of different ionization potentials exist in regions of different
electron densities.

In contrast to the post-shock case, here the timescales $\Delta t$
are determined by the ionization rates and may vary by an order of
magnitude or more from ion to ion.

From the SD-200 model, we adopt the average ionization potential as
a function of time and compute the ionization state distribution of
all elements with the simple approximation that this average
ionization potential is equal for all of them. We also take the
temperature profile from the model, which has some effect on the
resulting emissivities.
Fortunately, precise temperature values have only a minor effect on
the obtained abundances, since the infrared line emissivities depend
weakly on $T_{\rm e}$ above $10^3$~K or so.

In the pre-shock ionization front of the SD-200 model, each
initially neutral atom is consequently ionized with maximum ionic
abundance of each ion of the order of 50-100\%.
Therefore, the ionic abundances obtained in Table~\ref{TabLinesPre}
should be quite close to the elemental abundances, provided that the
$\Delta t$ values are estimated correctly and the pre-shock PIR
provides the major contribution to the line intensity%
\footnote{%
Abundance of Ar$^{4+}$ estimated from [\ion{Ar}{v}] lines is
much lower than the Ar elemental abundance,
presumably because only some fraction of Ar atoms is ionized as far as
Ar$^{4+}$ in the pre-shock region.}.

\bigskip

As a test case, we also included in Table~\ref{TabLinesPre} the
similarly computed O$^{2+}$ abundance from the [\ion{O}{iii}]
88.36~$\mu$m line.
The result is very close to the expected value of one (even too
close, given our uncertainties of about a factor of two),
illustrating the reliability of our method.

\begin{figure}
\begin{center}
\centerline{
    \rotatebox{270}{
        \includegraphics[height=0.95\linewidth]{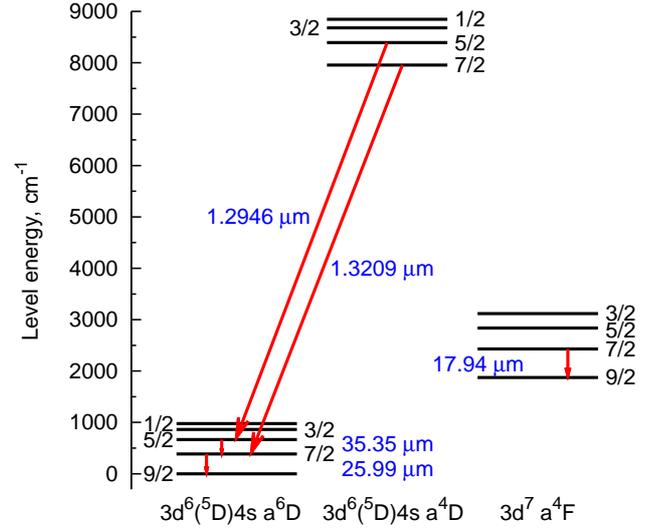}
                   }
           }
\caption{Energy diagram of the lowest levels of Fe$^+$ ion with
         transitions used in our analysis and corresponding line
         wavelengths. The electron configurations are given below
         the diagram; numbers besides the levels denote total angular
         momentum $J$. Data from \protect\citet{NISTASD}.
         }
   \label{FigFeII}
  \end{center}
\end{figure}

Although there are several [\ion{Fe}{ii}] spectral lines in the
infrared spectral range probed by \emph{Spitzer} that originate in
the low-lying electronic configurations, the two most promising are
blended with other stronger lines (one at 25.99~$\mu$m is blended
with the [\ion{O}{iv}] line at 25.91~$\mu$m and one at 35.35~$\mu$m
is blended with [\ion{Si}{ii}] line at 34.81~$\mu$m).
Therefore in the \emph{Spitzer} wavelength range, we used for our
analysis only the [\ion{Fe}{ii}] line at 17.94~$\mu$m (see
Fig.~\ref{FigFeII} for the energy levels of Fe$^+$ corresponding to
these transitions).

From Table~\ref{TabLinesPre}, it is seen that the Ar$^+$, Si$^+$,
and Fe$^+$ abundances estimated assuming that their fine-structure
lines arise exclusively in the pre-shock region are much higher than
the corresponding elemental abundance values obtained from optical
and X-ray data \citep{CheKirAbund,CasAXMM2002}.
The possible solutions of this discrepancy are discussed in
Sect.~\ref{SecDiscPre}.

\section{Post-shock photoionized region}
\label{SecPIR}

\subsection{On the too high pre-shock intensities of some lines}
\label{SecDiscPre}

Abundances of Ne and S derived assuming that the fine-structure
lines of [\ion{Ne}{ii}] and [\ion{S}{iii}] are emitted in the
pre-shock photoionized region of the SD-200 model agree reasonably
well with the results of optical and X-ray observations.
Simultaneously, similarly derived abundances of other singly-charged
ions (\element[+]{Si}, \element[+]{Ar}, \element[+]{Fe}) are
unreasonably high. They are even much higher than their estimated
elemental abundances (compare with Table~\ref{TabAbund}).

Within the frame of the SD-200 model, it is difficult to interpret
these results and at the same time to comment on the absence of some
other infrared lines, such as [\ion{Fe}{iii}] line at 22.93~$\mu$m
and many others.

There are several possible solutions to this mismatch between the
model predictions and observations.

\begin{itemize}

\item
The inconsistency might result from the application of the SD-200
model, describing the plasma dominated by C, O, Ne, and Mg, to the
plasma, containing considerable amounts of S, Si, and Ar. However,
the total abundance of these elements with respect to oxygen reaches
only about 10\%, which should not significantly alter the shock
structure.

\item
Flux in the Si and Fe infrared lines might mostly originate in
physically separated Fe-dominated clouds.
This hypothesis is based on two observations:
 (a) in [\ion{Fe}{ii}] line observations Cas~A has a different morphology than in the
[\ion{O}{iv}] line \citep{CasAIRAC,CasAIRS}; and
 (b) X-ray Chandra
observations indicate that hot Si- and Fe-dominated
blobs exist in Cas~A.\\
However, the Cas~A morphology in the Fe~K$\alpha$ X-ray line also
has a strongly different morphology from that in the [\ion{Fe}{ii}]
17.94~$\mu$m line weakening these arguments.
This hypothesis could be tested by observations of the Cas~A ejecta
with higher spectral resolution, which would allow one to determine
whether [\ion{Fe}{ii}] and other infrared lines from the same knots
have the same peculiar velocities.

\item
The [\ion{Si}{ii}], [\ion{Ar}{ii}], and [\ion{Fe}{ii}] lines might
originate in separate weakly-ionized photoionized region of high
emission measure. It might be situated either before the shock (the
``far precursor'' region of \citet{Itoh81a,Itoh81b}) or after it
(the post-shock PIR present in models of \citet{Itoh81a,Itoh81b} and
\citet{BorkowskiO}).
This hypothesis is discussed further below.

\end{itemize}

It is possible to distinguish between lines emitted by the pre-shock
and post-shock PIRs because of their strongly different electron
densities. For this purpose, we use additional constraints on this
line-emitting region that can be obtained from the \citet{CasANIR01}
observations of the near-infrared [\ion{Fe}{ii}] lines (see
Table~\ref{TabLinRat}).

The [\ion{Fe}{ii}] line ratio
$I$(1.3209~$\mu$m)/$I$(1.2946~$\mu$m$)<1$ (see Fig.~\ref{FigFeII}
for the transition diagram between the Fe$^+$ levels) strongly
indicates the electron density above $10^5$~cm$^{-3}$ at any
temperature, thus weakening the hypothesis of the ``far precursor''.
In the case of lower densities, this ratio is always higher,
reaching values of $10-20$ in the low-density limit because of
strong differences in the excitation rates of the respective
transition upper levels \citep{FeIIFS07}.

Another problem for the pre-shock ``far precursor'' hypothesis is
the observed ratio of the [\ion{Si}{ii}] 34.81~$\mu$m to the
[\ion{Fe}{ii}] 17.94~$\mu$m lines of about 2.
Both neutral and singly ionized Si and Fe atoms have similar
ionization potentials, therefore we might expect that their
ionization fractions are similar. From the line emissivities for the
low-density conditions and average abundance ratio of
$n$(Si)/$n$(Fe$)\approx10$ (Table~\ref{TabAbund}), the expected line
ratio is as high as 100.
We therefore conclude that the PIR contributing to the emission in
these lines is likely to extend \emph{after} the shock.

\subsection{Lines originating in the post-shock PIR}
\label{SecPIRLines}

All spectral lines of the neutral, singly-, and doubly-charged ions
originate in the pre-shock and the post-shock photoionized regions.
The strongly different densities of these regions (by a factor of
more than $10^3$) produces a much lower ionization in the denser
post-shock PIR, which is only partially ionized
\citep[e.g.,][]{BorkowskiO}.

In contrast, neutral and singly-ionized atoms are not abundant in the
pre-shock PIR, where they are rapidly ionized to higher degrees.
Estimates based on the SD-200 model confirm that for these atoms the
pre-shock PIR contribution to flux in most of their lines is below
10-30\%.

For example, the [\ion{S}{ii}] lines close to 6720~\AA\ are strongly
damped in the high-density medium of the post-shock PIR and are
obviously emitted before the shock. It provides the possibility of
evaluating directly the pre-shock contribution to the bright
near-infrared [\ion{S}{ii}] lines near 10\,300~\AA\ from the
low-density line ratios, which turns out to be about 10\%.

For our qualitative analysis of the post-shock photoionized region,
presented below, we use the lines listed in Table~\ref{TabPIRLines}.
There are also other lines that originate in the same region, but
their analysis does not yield additional conclusions.

\begin{table*}
\caption{%
Spectral lines used for analysis of the post-shock photoionized
region. See text for details. \label{TabPIRLines}}
\begin{center}
\begin{tabular}{ll|llc|l}
\hline\hline
 Spectrum       & $\lambda$    & $I/I($5007~\AA) & Source   & Observed? &
       Application  {\large\strut}\\
\hline
{}\ion{He}{i}   & 10830 \AA    &  0.006$^*$& GF01       & Yes?&
       He abundance {\large\strut} \\
{}[\ion{C}{i}]  & 8727 \AA     &  0.002    & HF96       & Yes &
       Post-shock PIR disruption  \\
{}[\ion{O}{i}]  
                & 6300 \AA     &   0.08    & HF96       & Yes &
       C abundance  \\
                & 63.19 $\mu$m & 0.07      & This paper & Yes &
       O ionization state, electron density,  \\
 & & & & & extent of the post-shock PIR \\
                 & 145.5 $\mu$m & $<$0.0024 & This paper & No  &
       Electron density \\
  \ion{O}{i}    & 7774 \AA     & 0.003     & HF96       & Yes &
       Extent of the post-shock PIR, O ionization state, \\
 & & & & & reference for line ratios \\
{}[\ion{Si}{ii}]& 34.81 $\mu$m & 0.07      & This paper & Yes &
       Extent of the post-shock PIR  \\
{}[\ion{S}{i}]  
                & 25.25 $\mu$m & $<$0.01   & This paper & No &
       S$^0$ abundance \\
{}[\ion{S}{ii}] &10287--10370 \AA& 0.11    & HF96, GF01 & Yes &
       Probable multi-$T$ structure of the post-shock PIR  \\
{}[\ion{Ar}{ii}]& 6.985 $\mu$m &  0.5      & This paper & Yes &
       Extent of the post-shock PIR, Ar ionization state\\
{}[\ion{Fe}{ii}]& 1.2946 $\mu$m&  0.001    & GF01       & Yes &
       Electron density \\
                & 1.3209 $\mu$m&  0.0007   & GF01       & Yes &
       Electron density and temperature \\
                & 17.94 $\mu$m &  0.03     & This paper & Yes &
       Electron temperature, extent of the post-shock PIR  \\
                & 25.99 $\mu$m &  $<$0.15  & This paper & No? &
       Electron temperature  \\
\hline
\multicolumn{6}{l}
    {$^*$ An upper limit to the line intensity intensity (blended lines)}\\
\end{tabular}
\end{center}
\emph{Note}. References: HF96 - \citet{HF96}, GF01 - \citet{CasANIR01}.
\end{table*}

\subsection{Charge exchange processes in post-shock PIR}
\label{SecChEx}

Atomic and ionic spectral lines in a rarefied partially photoionized
cold plasma are produced by
\begin{itemize}
  \item Collisional excitation from the ground state,
  \item Radiative recombination onto excited states, and
  \item Charge exchange reactions into excited states.
\end{itemize}

The charge exchange process at low temperatures is particularly
effective in our case of partially ionized plasma, as the atom-ion
charge exchange is not slowed by electrical repulsion of the
reactants.
Since the oxygen atom ionization potential is higher than those of
C, Si, S, and Fe, the charge exchange reactions may neutralize
oxygen simultaneously ionizing these elements.

It is interesting to note that the charge exchange process have
never before been accounted for in the theoretical FMK models. The
pure-oxygen FMK models \citep{Itoh81a,Itoh81b,BorkowskiO} do not
even mention charge exchange between, e.g., neutral and doubly
ionized oxygen as one of the relevant processes.
Based on the method of \citet{Dopita84}, the \citet{SD95}
computations do not include the charge exchange reactions because
the plasma temperature is too low for them to occur, according to
\citet{Hasted62}. However, their argument does not apply, since in
our case there are pseudo-crossings of intermittent molecular ion
potential energy curves.

We base our estimates of the rates of the relevant charge exchange
reactions on the simple procedure suggested by \cite{Peq86}
accounting for considerations of \citet{Stancil01}. Our estimates
are presented in Table~\ref{TabCX} with the spectral lines produced
by a charge exchange into excited states (so-called charge exchange
excitation, see, e.g., \citet{DalgarnoCX82,DalgarnoCX85}).

\begin{table*}
\caption{%
Charge exchange reaction rates relevant to the post-shock PIR in an
oxygen-dominated medium. \label{TabCX} }
\begin{center}
\begin{tabular}{l|lll|l}
\hline\hline
 Reaction
   & $\Delta E$, eV & $\delta E$, eV & $\alpha_{\rm CX}$, cm$^3$/s
   & Emitted lines; notes \\
\hline
O$^+$ + C$^0$ $\to$ O$^0$($^1$D) + C$^+$
   & 2.36 & 0.39 & $1\times10^{-9}$ &
   [\ion{O}{i}] 6300, 6363~\AA\\
O$^+$ + Si$^0$ $\to$ O$^0$ + Si$^+$(3s 3p$^2$ $^4$P)
   & 5.47 & 0.15 & $3\times10^{-9}$ &
   [\ion{Si}{ii}] 2329--2351~\AA\\
O$^+$ + S$^0$ $\to$ O$^0$ + S$^+$($^2$P)
   & 3.26 & 0.22 & $3\times10^{-9}$ &
   [\ion{S}{ii}] 4069, 4076, 6716, 6731, 10287--10370~\AA\\
O$^+$ + Fe$^0$ $\to$ O$^0$ + Fe$^+$ (excited)
   & 5.72 & many transitions & $1.7\times10^{-9}$ &
   multiple lines; rate from \citet{Rutherford72}\\
%
\hline
\end{tabular}
\end{center}
\emph{Notes}. It is assumed that the atoms and ions are in their
ground states (or excited fine-structure states) before the
charge-exchange reaction. $\Delta E$ denotes the difference in the
ionization potentials of atom and ion in their ground states,
$\delta E$ denotes the same difference for outgoing atom and ion in
specified states, and $\alpha_{\rm CX}$ is the charge exchange rate
estimate. Only excited state classification is shown.
\end{table*}

\begin{figure}
\begin{center}
\centerline{
    \rotatebox{270}{
        \includegraphics[height=0.95\linewidth]{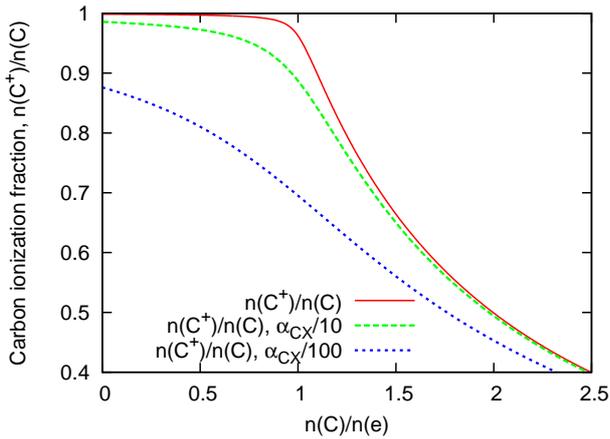}
                   }
           }
\caption{Ionization state of carbon in an oxygen-carbon partially
         photoionized plasma. Since carbon and oxygen photoionization and
         recombination rates are similar, the carbon ionization fraction is
         almost exclusively dependent on the ratio $n_{\rm C}/n_{\rm e}$.
         Dependencies on both total atomic density and carbon abundance
         at constant $n_{\rm e}$ are minor.
         The upper curve corresponds to the charge exchange rate given
         in Table~\ref{TabCX}, and the middle and lower curves to the rates
         artificially diminished by factors of 10 and 100.
         It is seen that the charge exchange processes result in carbon being ionized
         by more than 95-99\% when the number density of carbon atoms and ions
         is less than that of electrons.
         }
   \label{FigCX}
  \end{center}
\end{figure}

Sicne the charge exchange is by far the most rapid of collisional
processes, the ionization state of various elements in the
post-shock PIR is insensitive to the charge exchange rate values.
Taking these rates lower by one order of magnitude produces only an
insignificant change in the ionization states (see Fig.~\ref{FigCX}
for example).
Roughly, the number of ionized oxygen atoms decreases by the total
number of C, Si, S, and Fe atoms, provided that $n_{\rm O}>n_{\rm C}
+ n_{\rm Si} + n_{\rm S} + n_{\rm Fe}$.

The charge exchange reactions neutralizing singly charged ions of
neon and argon in collisions with atoms of oxygen is estimated to
be very slow%
\footnote{We are grateful to the anonymous referee for pointing out
the source of the Ne and Ar charge exchange rates with oxygen.} %
(of the order of $1\times10^{-13}$~cm$^3$/s,
\citet{LiuDalgarno1996}), therefore the argon and neon ionization
state in the post-shock PIR is controlled by ionization and
recombination processes. Charge exchange of Ne and Ar ions with
other neutral atoms (C, Si, S, Fe) is not important since these
elements are mostly ionized (see above).
This qualitatively explains the strong observed [\ion{O}{i}] line at
63.19~$\mu$m and the absence of fine-structure [\ion{Si}{i}],
[\ion{S}{i}], and [\ion{Fe}{i}] infrared lines.

The weakness of the neutral oxygen recombination line at 7774~\AA\
is also indirect proof of the charge exchange process operating in
the post-shock PIR
\footnote{We are grateful to the anonymous referee for pointing out
the usefulness of the \ion{O}{i} 7774~\AA\ spectral line.}. %
The 7774~\AA\ line is traditionally used to estimate the extent of
this region \citep{Itoh86}, as the line intensity is proportional to
the amount of oxygen recombination (in the absence of charge
exchange, also to ionization) events, which depends on the optical
depth of the post-shock PIR.
At the post-shock PIR temperatures, each radiative recombination of
a singly ionized oxygen atom is followed by the 7774~\AA\ line
emission with a probability of about $p=0.4$ \citep{Peq86}.

However, since ionized oxygen in the post-shock PIR may be
neutralized by both recombination and charge exchange, the 7774~\AA\
line weakens, and its flux alone does not provide enough data to
determine the optical depth of the post-shock PIR.
To obtain full information about the number of recombinations in the
post-shock PIR plasma, deeper observations are needed that would
also detect optical and near-infrared recombination lines of other
elements (e.g., \ion{C}{i} line at 10695~\AA, \ion{S}{i} line at
9237~\AA, etc.).

However, a lot of useful information may be extracted from the flux
of the 7774~\AA\ recombination line. For example, one can roughly
estimate the oxygen ionization from the ratio of its flux to the
total flux from the post-shock PIR.

If plasma contains only oxygen, then the number of O recombination
events will be equal to the number of absorbed ionizing photons. The
presence of other elements with similar ionization cross-sections
and recombination rates will decrease the number of oxygen
recombinations by a factor $n_{\rm O}/n_{\rm t}$, if $n_{\rm t}$
remains constant and no charge exchange reactions occur.
In this case, the ratio of the 7774~\AA\ line flux to the total flux
emitted from the post-shock PIR cannot be smaller than
$$
  \frac{I(7774~\mbox{\AA})}{\sum I}
      = \frac{p\, h\nu_{\rm 7774}}{h\nu_{\rm ioniz}}
        \frac{n_{\rm O}}{n_{\rm t}}
      = \frac{0.4 \times 1.6 \mbox{ eV}}{27 \mbox{ eV}}
        \frac{n_{\rm O}}{n_{\rm t}}
      = 0.024
        \frac{n_{\rm O}}{n_{\rm t}},
$$
where $h\nu_{\rm ioniz}$ is the average ionizing photon energy. It
is estimated to be
$$
  h\nu_{\rm ioniz}
  = \frac12 \times \frac{H_{\rm sh}}{N_{\rm ph}}
  = \frac12 \times \frac
{(5/2)k_{\rm B}T_{\rm sh} \times (n_{\rm e}/n_{\rm t}+1) n_{\rm t}/n_{\rm O}}
       {N_{\rm ph}},
$$
where $T_{\rm sh}$ and $n_{\rm e}/n_{\rm t}$ are the temperature and
the electron-to-atom ratio at the shock front, respectively, $H_{\rm
sh}\approx2250$~eV is the enthalpy per oxygen atom at the shock
front (the cooling post-shock plasma should be isobaric most of the
temperature range), and $N_{\rm ph}\approx42$ is the average number
of emitted ionizing photons per oxygen atom (its value being derived
from the ionization parameter $Q$ of \citet{SD95})  in the SD-200
model. The factor of $1/2$ is the fraction of downstream photons.


However, the observed value of this 7774~\AA\ line ratio is at least
a factor of three lower (its ratio with respect to the FS line of
[\ion{Ar}{ii}] alone is about 0.007, see Tables~\ref{TabPIRLines}
and~\ref{TabAIR}).
We note that this ratio of 1/3 is an absolute upper limit. The ratio
of the expected 7774~\AA\ line flux to the observed one may be much
higher, since part of the much stronger dust continuum emission (of
flux 10-40 times higher than that in the 5007~\AA\ line) may
originate in the post-shock PIR.

From this, we infer that at least two thirds of the ionized oxygen
atoms recombine by means of charge exchange. Assuming additionally
that
\begin{itemize}
\item this results in a complete ionization of atoms of other elements
connected to oxygen by charge exchange reactions (see above),
\item the recombination rates of all singly ionized atoms are similar, and
\item the average ionization degree of plasma is about 40\% \citep{BorkowskiO},
\end{itemize}
we find that the total number of elements related to oxygen by
charge exchange reactions constitutes about $2/3$ of the number
density of oxygen itself (i.e., $n_{\rm O}/n_{\rm t}\approx 0.6$)
and the oxygen ionization $n({\rm O}^+)/n({\rm O})$ is not higher
than 0.10--0.15.


\subsection{The post-shock PIR temperature and density}
\label{SecPIRPar}

Once the physical situation has been established, we now determine
the conditions in the post-shock photoionized region.

The lower limit to the temperature of about 1000~K is obtained from
the requirement that the [\ion{Fe}{ii}] 25.99~$\mu$m line emitted by
this region should not be too bright (an observational constraint is
$I$(25.99~$\mu$m)/$I$(17.94~$\mu$m$)<5$, see
Sect.~\ref{SecSpitzerO}).

The very low value of the near-infrared [\ion{Fe}{ii}] 1.3209~$\mu$m
line ratio to the 17.94~$\mu$m line corresponds to $T_{\rm e}
\approx 1700$~K in the high-density limit (the high electron density
follows from the [\ion{O}{i}] fine-structure line ratio, see
Sect.~\ref{SecOI}).
This [\ion{Fe}{ii}] line ratio is very sensitive to the temperature
because of the different excitation potentials of the transitions
upper levels and possibly provides only an upper limit to the
temperature with the near-infrared 1.3209~$\mu$m line partially
arising in some smaller and hotter region.

Assuming in our analysis the PIR temperature of 1500~K, we can
obtain stronger constraints on the electron density from the limit
on the FIR [\ion{O}{i}] line ratio, assuming that these lines
originate in the post-shock PIR, as discussed in
Sect.~\ref{SecPIRSize}.

In this case, the FIR [\ion{O}{i}] line ratio corresponds to $n_{\rm
e}\ga 5\times10^5$~cm$^{-3}$.
We assume this value for the electron density and twice as large a
value $n_{\rm t}=1\times10^6$~cm$^{-3}$ for the total atom and ion
number density (average ionization of about 40-50\% is obtained by
\citet{BorkowskiO} and should be approximately correct, since
recombination rates of all singly charged ions are similar).

We note that based on the assumption of constant post-shock gas
pressure this $n_{\rm t}$ value corresponds to the pre-shock number
density of about one hundred atoms per cm$^{3}$, implying that the
magnetic field does not significantly modify the post-shock plasma
compression.

\subsection{The post-shock PIR ionization and thickness}
\label{SecPIRSize}

The ratio of the neutral oxygen lines
$I$(7774~\AA)/$I$(63.19~$\mu$m$) \approx 0.045$ may be used to infer
the oxygen ionization degree, which is found to be $n({\rm
O}^+)/n({\rm O}^0)\approx 0.06$ for our assumed electron density.
This estimate is precise up to a factor of two if $n_{\rm e}$ is in
the range between $1\times10^5$ and $1\times10^6$~cm$^{-3}$ and only
weakly depends on the region temperature.


The value of the oxygen ionization degree is in reasonable
accordance with the estimates of Sect.~\ref{SecChEx}. This is also a
reason for us to assume that most of the 63.19~$\mu$m emission
originates in the FMKs, and not in some more diffuse regions around
them (remember that this line was observed by ISO, which had poor
angular resolution).
In its turn, this allows one to use the [\ion{O}{i}] 63.19~$\mu$m
line intensity to estimate the extent of the post-shock PIR,
assuming that it is homogeneous.

To estimate the post-shock PIR thickness (see Fig.~\ref{FigStruct}),
we can also use the fine-structure lines of [\ion{Si}{ii}],
[\ion{Ar}{ii}] and [\ion{Fe}{ii}], as well as the \ion{O}{i}
7774~\AA\ recombination line. These estimates are summarized in
Table~\ref{TabPIRSize}.

\begin{table}
\caption{%
Estimates of the post-shock photoionized region thickness $l_{\rm
PSPIR}$ from several fine-structure line ratios to the 5007~\AA\
line. \label{TabPIRSize}}
\begin{center}
\begin{tabular}{l|ll|l}
\hline\hline
Spectral line & $\varepsilon$, erg cm$^3$/s & $n($ion$)/n($elem$)$
          & $l_{\rm PSPIR}$, cm \\
\hline
{}[\ion{O}{i}]   63.19~$\mu$m & $1.2\times10^{-24}$ & 0.94& $1.2\times10^{11}$ \\
{}[\ion{Si}{ii}] 34.81~$\mu$m & $1.3\times10^{-23}$ & 1.0 & $2.1\times10^{11}$ \\
{}[\ion{Ar}{ii}] 6.985~$\mu$m & $2.0\times10^{-21}$ & 0.4 & $2.0\times10^{11}$ \\
{}[\ion{Fe}{ii}] 17.94~$\mu$m & $4.0\times10^{-23}$ & 1.0 & $3.0\times10^{11}$ \\
   \ion{O}{i}    7774~\AA     & $1.3\times10^{-24}$ & 0.06& $8.0\times10^{10}$ \\
\hline
\end{tabular}
\end{center}
\emph{Notes}. Estimates made for post-shock PIR electron density
$n_{\rm e}=5\times10^5$~cm$^{-3}$ and temperature $T_{\rm
e}=1500$~K. Oxygen abundance have been taken to be $n_{\rm O}/n_{\rm
t}=0.6$ (see Sect.~\ref{SecChEx}), other abundances as in
Table~\ref{TabAbund}.
\end{table}

%
Closer agreement between the region sizes obtained from the
[\ion{O}{i}] and [\ion{Fe}{ii}] lines may be achieved if the region
temperature is assumed to be about 1800--2000~K (then the
[\ion{Fe}{ii}] line emissivity will increase decreasing its region
size estimate to $(2.0-2.2)\times10^{11}$~cm).

The estimate from the 7774~\AA\ line also becomes closer to the
value derived from the [\ion{O}{i}] line at these slightly higher
temperatures, becoming about $9.3\times10^{10}$~cm.
Alternatively, decreasing the oxygen ionization degree increases the
post-shock PIR size estimate obtained from this line.

It is worth noting that values of the post-shock PIR size in
Table~\ref{TabPIRSize} are only about factor of about two smaller
than the value of $3\times10^{11}$~cm predicted by
\citet{BorkowskiO}.
This observational evidence therefore does not require significant
truncation of the post-shock region, proposed by \citet{Itoh86}.
However, other spectroscopic signatures, discussed in
Sect.~\ref{SecTrunc}, show that some truncation of the post-shock
PIR seems to be present.

\subsection{Amount of carbon in the FMKs}
\label{SecPIRC}

The [\ion{O}{i}] 6300~\AA\ line originates in neither the pre-shock
PIR, nor the post-shock cooling region, as shown by \citet{SD95},
therefore its source is located in the post-shock PIR.
Since it is a factor of 25 stronger than the \ion{O}{i} 7774~\AA\
recombination line, we can conclude that recombination provides only
a minor contribution to the line flux.
It may be produced by collisional excitation, but then the
corresponding temperature from the $I$(6300~\AA$)/I$(63.19~$\mu$m)
line ratio of about unity corresponds to temperatures of at least
3000~K.

However, the 6300~\AA\ line may be produced by a charge
exchange reaction with carbon%
\footnote{Note that the charge exchange with other elements do not
produce the 6300~\AA\ line; carbon is unique in this respect.} %
(see Table~\ref{TabCX}).
As this reaction is very fast, almost all the carbon is ionized and
every carbon recombination will be immediately followed by the
charge exchange reaction with an oxygen ion. Then, with a
probability of about 72\% (accounting for collisional transitions
downwards), oxygen emits the 6300~\AA\ line. A rather strong
observed 6300~\AA\ line infers a considerably high abundance of
carbon in the FMK plasma.

Specifically, from this line ratio to the 7774~\AA\ line, one can
conclude that the carbon abundance is
$$
n_{\rm C}/n_{\rm O} \approx 0.5,
$$
assuming that all carbon is ionized.

This result is only weakly dependent on the post-shock PIR
temperature, as long as it stays below 3000~K, where collisional
excitation of 6300~\AA\ begins to play a role, and the electron
density, as long as it is lower than about $5\times10^6$~cm$^{-3}$,
where collisional de-excitation in neutral oxygen starts to diminish
the 6300~\AA\ line emissivity. However, estimated carbon abundance
is proportional to the assumed oxygen ionization, taken here to be
0.06 (see Sect.~\ref{SecPIRSize}) and is therefore uncertain by up
to a factor of two.

\subsection{Sulfur in the post-shock PIR}
\label{SecPIRS}



From the non-detection of the [\ion{S}{i}] fine structure line at
25.25~$\mu$m, we can define an upper limit to the neutral sulfur
abundance in the post-shock PIR. For this purpose, we compare the
[\ion{S}{i}] 25.25~$\mu$m line with the [\ion{O}{i}] 63.19~$\mu$m
line having similar dependence on temperature and density.
The upper limit on their ratio %
$I$(25.25~$\mu$m$)/I$(63.19~$\mu$m$)<0.14$ %
corresponds to the abundance ratio of %
$n({\rm S}^0)/n({\rm O}^0)<0.004$, showing that sulfur is more than
90\% ionized.


We note that in the framework of our model it is difficult to
understand the brightness of the [\ion{S}{ii}] lines near
10\,300~\AA.
As discussed in Sect.~\ref{SecPIRLines}, they are emitted after the
shock. However, in our simple single-temperature model none of the
line production mechanisms -- radiative recombination, charge
exchange with oxygen (see Table~\ref{TabCX}), or collisional
excitation -- can explain the high brightness of these lines.


\subsection{Helium in the optical FMKs}
\label{SecPIRHe}

X-ray data infers that a large amount of elements with $Z<8$ are
present in the hot Cas~A ejecta \citep{ChandraCasA2000}. Obviously,
these elements are hydrogen, helium or carbon.
%
\citet{DeweyO} even suggested that hot X-ray emitting plasma
consists of more than 95\% He. This is also consistent with the
suggestion by \citet{SD95} that a hot plasma between the optical
FMKs would be much brighter in X-rays than observed, if it consisted
mostly of oxygen.

The amount of helium in the FMK plasma may be constrained, e.g.,
from the upper limit to its well-known near-infrared recombination
line at 10\,830~\AA.
Each singly charged helium ion recombination with a probability of
about 14\% results in the emission of this line \citep{RRC91}.
Therefore, the total recombination of helium after the shock will
result in $0.14 n({\rm He})/n({\rm O})$ photons per oxygen atom.

Comparing this with the SD-200 model 5007~\AA\ line intensity of
about 7~photons per oxygen atom, we would expect the
$I$(10830~\AA$)/I$(5007~\AA$)\approx0.02 n({\rm He})/n({\rm O})$.
The observed line ratio upper limit of 0.006 thus corresponds to the
helium abundance upper limit of about $n({\rm He})/n({\rm O})\la0.3$.
This result is consistent with the \citet{CheKirCasA} and
\citet{Peimbert71} estimates, but is $5-10$ times more restrictive
because of more certain theoretical model of the post-shock PIR.

This value is probably underestimated by up to a factor of two as
helium will still be partially ionized in the post-shock PIR and may
not recombine fully if this region is disrupted (see below).
Nevertheless, it is obvious that the helium abundance of optical
FMKs is not very high, implying that optical and X-ray emitting
ejecta indeed have different chemical abundances.

\subsection{Evidence of post-shock PIR disruption from spectral lines}
\label{SecTrunc}

Post-shock region disruption have been observed in both numerical
simulations of the shock interaction with the dense interstellar
clouds in supernova remnants \citep{CloudShockSim05} and in laser
experiments emulating this interaction \citep{CloudCrushExp03}.

As shown by \citet{Itoh86}, there are spectroscopic signatures that
allow one to determine if the post-shock PIR is disrupted.
Above, we have argued that the original diagnostics of
\citet{Itoh86} -- the \ion{O}{i} 7774~\AA\ line -- is not applicable
because of the charge exchange between oxygen and other elements,
and the poorly constrained typical chemical composition of the FMKs.

Instead we consider other spectral lines that are sensitive to the
final recombination in the post-shock PIR.
Indeed, if this region is present, then each ion in the post-shock
PIR recombines at least once, giving a lower limit to the line
intensities due to recombination.
The most sensitive to the presence of final recombination are
high-excitation lines of atoms that should be completely ionized due
to charge exchange with oxygen, i.e., ones that should not be
emitted in any other region of a FMK.

As an example, we consider the observed [\ion{C}{i}] line at
8727~\AA.
It is emitted with probability of roughly 10\%, when a carbon ion
recombines.
Thus, there should be at least 0.05~photons in this line per every
oxygen atom (using the carbon abundance from Sect.~\ref{SecPIRC}).
Comparing this with the SD-200 model 5007~\AA\ line intensity of
about 7~photons per oxygen atom, we would expect to measure
$I$(8727~\AA$)/I$(5007~\AA$)\approx0.007$.
The observed 8727~\AA\ line is a factor of three less intense,
implying that not all carbon recombines after the shock.

Unfortunately, this result, which confirms the post-shock PIR
disruption, is not very robust, since the carbon abundance value is
not very tightly constrained.
However, similar diagnostics may be performed with this and other
lines using future more sensitive observations. These observations
will show us the maximum extent to which each atomic species
recombines in the FMKs after the reverse shock.

We note that another argument exists for the post-shock PIR
truncation, which is based on energy considerations%
\footnote{This argument was pointed out by the
anonymous referee.}.%
The energy of the post-shock PIR line emission per oxygen atom is a
factor 5--10 less than the energy that should be deposited if the
emission were optically thick to the post-shock cooling plasma
emission. The most obvious solution to this problem is post-shock
PIR truncation.

However, another possibility is that some part of the cooling plasma
energy is transferred to dust heating inside the post-shock cooling
region or post-shock PIR and is then emitted in the infrared dust
continuum, which in total emits a factor of 12 more energy than the
lines from the post-shock PIR (see Sect.~\ref{SecEnergy}).
Most probably, a combination of these two factors (disruption and
continuum emission) explains the apparent difference between the
expected and observed post-shock PIR emission-line fluxes.

\subsection{Effects of the dust on the post-shock PIR structure}

Using the data from \citet{CasAIRS}, it is straightforward to
estimate the dust mass in each pixel of the \emph{Spitzer} map of
the Cas~A. Its comparison with the optical image of this supernova
remnant allows us to determine the dust-to-gas mass ratio in the
bright knots from the [\ion{O}{iii}] line flux predicted by the
SD-200 model.

If the emitting dust with temperature $T_{\rm d}\approx 100$~K
\citep{CasAIRS} is distributed over the entire volume of the knot,
then the dust-to-gas mass ratio is of the order of 2-5\%. If, on the
other hand, the dust emits only in the line-emitting regions, then
the dust-to-gas mass ratio is of the order of unity.

If the latter is true, then the current shock models do not apply
and construction of an entirely different model is required to
describe the shock structure of the Cas~A knots.
This analysis is outside the scope of this paper.

\section{Recombination lines in the infrared range}
\label{SecIRL}

The metal recombination lines (RLs) are another good tracer of the
cold ionized plasma. Their emissivities as functions of temperature
have no exponential cutoff at low $T_{\rm e}$ and increase with
decreasing temperature approximately as $T_{\rm e}^{-1}$ at $T_{\rm
e}<10^4$~K.
The line emissivities are only weakly dependent on density.

However, their emissivities are $4-6$ orders of magnitude less than
those of the fine-structure infrared lines at the low-density limit.
The infrared RL emissivity ratios to the fine-structure lines
increase with electron density due to collisional effects that
diminish the fine-structure line emissivities as $n_{\rm e}^{-1}$
starting from some critical density.
At post-shock electron densities of $10^6$~cm$^{-3}$, the emissivity
ratios are much higher \citep{CasAORL}, but the infrared RLs are
nevertheless still far less intense than the fine-structure lines.

As two examples, we consider infrared RLs near 10 and 60~$\mu$m.
The brightest regions emitting RLs are the post-shock cooling and
photoionized regions (due to their high emission measures $n_e^2 l$)
and cold region between the photoionization front and the shock (due
to its extremely low temperature and large extent).

The \ion{O}{i} 5$\alpha$ recombination line at 7.45~$\mu$m
originates in the post-shock photoionized region and its strongest
component emissivity of $5\times10^{-27}$~erg cm$^{3}$/s is about
200 times less than that of the \ion{O}{i} 7774~\AA\ line.
The expected recombination line flux is therefore $3\times10^4$
times less than that of the [\ion{Ar}{ii}] 6.985~$\mu$m line, or
about 0.5\% of the background continuum emission. However, other
elements will also produce recombination lines at the same
wavelength (e.g., ionized carbon is almost an order of magnitude
more abundant than ionized oxygen, see Sect.~\ref{SecPIRC}) and the
total $5\alpha$ line flux may be a factor of 10-20 more.
Estimates show that this limit is achievable with \emph{Spitzer}
with exposure times of a few hours.

Another example is a \ion{O}{i} 11$\alpha$ RL at 69.03~$\mu$m.
Its emissivity is approximately 5000~times less than that of nearby
[\ion{O}{i}] fine-structure line at 63.19~$\mu$m. Expected
intensity, if summing with 11$\alpha$ lines of other ions, is about
$5\times10^{-15}$~erg/cm$^2$/s from the ISO region \#2.
This is approximately equal to the expected 5$\sigma$ 1-hour point
source sensitivity of the Herschel PACS instrument~\citep{PACS06}.
Since the Cas~A is not a point source at the Herschel resolution
(PACS has a 10'' pixel size and most of the Cas~A emission is
contained within 4-8 PACS pixels), one needs at least a few hours to
achieve the 5$\sigma$ detection of this recombination line.

The recombination lines of highly-charged oxygen ions from the
post-shock cooling region are expected to be about an order of
magnitude dimmer than the \ion{O}{i} RLs because of the lower
emission measure in the region emitting these lines.

The metal recombination line observations in the mid- and far-infrared
are generally more difficult than in the optical and near-infrared due to
lower line emissivities and higher background continuum emission.
They are also unable to provide information about the ion producing
the spectral line, but only about its electronic charge.
However, these observations would allow us to determine abundances
of ions residing in cold regions that do not produce fine-structure
lines. Special emphasis should be placed on the singly-charged ion
lines, since their recombination lines in the optical and
near-infrared ranges are divided into many weak components, which
makes them more difficult to detect \citep{CasAORL}.

\section{Conclusions}
\label{SecConclusions}

We have performed an analysis of the supernova remnant Cas~A
fast-moving knot infrared line intensities, by comparing
observations with the predictions of various theoretical models,
which describe the FMK emission originating from the reverse shock
interaction with the pure-oxygen or oxygen-dominated clouds.
For this comparison, we have analyzed archival observational data
acquired by the ISO and \emph{Spitzer} observatories.
We conclude that accounting for the electron conductivity is
essential to reproducing the observed line ratios. This is the
reason why the \citet{BorkowskiO} model BS-DC provides the most
accurate description of the observed oxygen line relative fluxes,
although it is only precise up to a factor of several (the
[\ion{O}{i}] lines are overpredicted partially because of the too
low model pre-shock density).

This emphasizes the need for more sophisticated models describing
shock propagation in the oxygen-dominated plasma.
In this article, many line flux ratios are derived that will help us
to construct these future theoretical models.
Future far-infrared and optical observations of higher sensitivity
and angular resolution will allow one to obtain far more information
about the fast-moving knots once the models are constructed.

Analysis of the infrared lines of O, Ne, Si, S, Ar, and Fe ions have
confirmed the existence of three regions contributing to the
infrared and optical line emission:
\begin{itemize}
\item Lines of the intermediate- and highly-charged ions that arise in the
      post-shock rapidly cooling region
   (e.g., [\ion{Ne}{v}], [\ion{Ar}{v}], [\ion{Si}{vi}], [\ion{Si}{x}] lines).
\item Singly-charged ions that radiate mostly from the post-shock
      photo-ionized region
   (e.g., [\ion{O}{i}], [\ion{Si}{ii}], [\ion{Ar}{ii}], [\ion{Fe}{ii}] lines).
\item Weakly- and intermediate-charged ion lines originate
      in the pre-shock photoionized region as well.
      It gives major contribution to the emission of lines of
      [\ion{Ne}{ii}], [\ion{O}{iii}], [\ion{S}{iii}], [\ion{Ar}{v}], etc.
\end{itemize}

An analysis of the FIR observations jointly with the \citet{HF96}
and \citet{CasANIR01} optical and near-infrared data has allowed us
to construct a consistent model of the post-shock photoionized
region. This became possible by accounting for the charge exchange
process that diminishes oxygen ionization in the post-shock PIR.
Observed high ($>0.9$) ionization of sulfur and low ($\approx0.06$)
ionization of oxygen confirms the importance of the charge exchange
reactions in this region.

From the observed line flux ratios, it then follows that the
post-shock PIR has a temperature $T_{\rm e}\approx (1500-2000)$~K,
electron and ion densities of $n_{\rm e}\approx
5\times10^5$~cm$^{-3}$ and $n_{\rm t}\approx 1\times10^6$~cm$^{-3}$
and is approximately $(1-2)\times10^{11}$~cm thick, assuming that
oxygen constitutes about 70\% of all atoms in the FMK plasma.
Given the uncertainties of the model, this result is consistent with
the thickness predicted in the \citet{BorkowskiO} model.

We have constrained the helium abundance to be $n({\rm He})/n({\rm
O}) < (0.3-0.6)$ and the carbon abundance to be about $n({\rm
C})/n({\rm O}) \approx 0.5$.

However, we recall that the obtained values of elemental abundances
and ionization fractions in the post-shock PIR are affected by
unknown biases, as our assumptions (plane-shock, single-temperature
model) clearly do not adequately represent the structure of this
region.

From the comparison of observed infrared line intensities with the
theoretically expected values, we conclude that
\begin{itemize}
\item The theoretical models correctly describe the general
      structure of the FMKs;
\item Accounting for the electron conductivity brings the
      BS-DC model into far closer agreement with the observed
      oxygen FIR line ratios to the [\ion{O}{iii}] 5007~\AA\ line than the
      BS-F and SD-200 models;
\item Accounting for the emission from the photoionized region
      before the shock front is essential to reproducing the FIR
      [\ion{O}{iii}] line ratio;
\item The FIR [\ion{O}{i}] line intensities in the \citet{Itoh81b}
      and \citet{BorkowskiO} models are partially overestimated
      because of the too low pre-shock atom number densities;
\item The pre-shock atom number density is at least 100~cm$^{-3}$
      and may be as high as 300~cm$^{-3}$;
\item The high density of the post-shock PIR suggests that there is
      no significant magnetic-field pressure in this region;
\item The post-shock photoionized region truncation, as suggested
      by \citet{Itoh86}, may be needed to explain the weakness of several
      lines produced by the final plasma recombination.
\end{itemize}

The infrared metal recombination lines are shown to be detectable by
the planned far-infrared instruments and useful for the derivation
of the plasma properties.

The high brightness of the [\ion{S}{ii}] lines close to 10\,300~\AA\
remains unexplained by our single-temperature model of the
post-shock PIR.

\begin{acknowledgements}

DD is thankful to Mike Revnivtsev for many useful advices
on data processing.
%

The version of the ISO data presented in this paper correspond
to the Highly Processed Data Product (HPDP) set called
'Uniformly processed LWS L01 spectra'
by C.~Lloyd, M.~Lerate and T.~Grundy,
available for public use in the ISO Data Archive,
{\tt http://www.iso.vilspa.esa.es/ida/}.

Some of the data presented in this paper were obtained
from the Multimission Archive at the Space Telescope Science Institute
(MAST).
They are based on observations made with the
NASA/ESA Hubble Space Telescope,
obtained from the Data Archive
at the Space Telescope Science Institute,
which is operated
by the Association of Universities for Research in Astronomy, Inc.,
under NASA contract NAS 5-26555.
These observations are associated with program \# 10286.

Part of this work is based on observations made with
the \emph{Spitzer} Space Telescope, which is operated by
the Jet Propulsion Laboratory, California Institute of Technology
under a contract with NASA.
This research made use of Tiny Tim/Spitzer, developed by John Krist
for the Spitzer Science Center.
The Center is managed by the California Institute of Technology
under a contract with NASA.

CHIANTI is a collaborative project involving
the NRL (USA), RAL (UK), MSSL (UK), the Universities of Florence (Italy)
and Cambridge (UK), and George Mason University (USA).

\end{acknowledgements}



\begin{appendix}

\section{\emph{Spitzer} data cube PSF size estimate}
\label{AppPSF}

As described by \citet{cubism}, the mathematical transformations
performed by the CUBISM software change the point spread function
(PSF) of the final data cube away from the initial PSF of the
\emph{Spitzer} IRS.
This effect should be more pronounced in images with only a few
pixels per PSF width, as constructed by the \emph{Spitzer} IRS data
cubes in the second order of each spectrograph module.

To determine the PSF changes induced by the data cube
reconstruction, we fit a two-dimensional Gaussian function to the
shape of a point-like source \object{2MASS 23233176+5853204}
situated in the map of the SL2 module.
This fit indicated that (here $x$ is the coordinate along the
individual slits and $y$ -- across them):
\begin{itemize}
\item The source $x$ centroid oscillates with wavelength
with an amplitude of about 0.3~pixels. The $y$ centroid remains
constant to within 0.05~pixels.
\item The PSFs in both the $x$ and $y$ directions are larger than the
\emph{Spitzer} IRS true PSF, computed, e.g., by the {\tt stinytim}
software. The full width at half maximum (FWHM) in the $x$ and $y$
directions of the spectral maps are greater than the IRS true PSF by
0.55~pixels and 0.9~pixels.
\end{itemize}

The presence of these features is inferred by the CUBISM algorithm
description in \citet{cubism}.
Although the values of the increase in PSF dimensions may seem
insignificant at first, they are often comparable with the extent of
the IRS true PSF (e.g., at 26~$\mu$m, the initial FWHM of the PSF is
equal to the 1.2 LL1 module pixel).

Not accounting for the described increase in the PSF size because of
processing by the CUBISM software in our case would result in the
\emph{Spitzer} maps being seemingly more diffuse than e.g., optical
maps smoothed to the IRS angular resolution.

Although we have measured these effects only in the SL2 module, we
assume that the same PSF broadening \emph{in pixels} occurs in other
modules as well. This conclusion is qualitatively confirmed by
visual comparison of optical maps smoothed to the corresponding
resolution of the data cube spectral line maps.

\section{Derivation of Eq. (\protect\ref{EqfPR})}
\label{AppfPR}

For the approximation of two homogeneous emitting regions 1 and 2,
the observed line $a$ and $b$ flux ratio $R\equiv I_a/I_b$ may be
expressed as
$$
R  \equiv \frac{I_a}{I_b}
   = \frac{I_{a,1} + I_{a,2}}{I_{b,1}+I_{b,2}}
   = \frac{I_{a,1}}{I_{b,1}} \frac{I_{b,1}}{I_{b,1}+I_{b,2}} +
     \frac{I_{a,2}}{I_{b,2}}
          \left( 1 - \frac{I_{b,1}}{I_{b,1}+I_{b,2}} \right).
$$
Denoting the line flux ratios arising in regions 1 and 2 as
$R_1 \equiv I_{a,1}/I_{b,1}$ and $R_2 \equiv I_{a,2}/I_{b,2}$,
and introducing the fraction of the total line $b$ emission arising
in the region 1 as
$$
f_{b,1} \equiv \frac{I_{b,1}}{I_{b,1}+I_{b,2}},
$$
we obtain
$$
  R = R_1 f_{b,1} + R_2 \left( 1 - f_{b,1} \right).
$$
Expressing $f_{b,1}$ from this linear relation, we finally derive
Eq.~(\ref{EqfPR}):
$$
  f_{b,1} = \frac{R_2-R}{R_2-R_1}.
$$

\end{appendix}

\end{document}